\pdfoutput=1

\documentclass[11pt,twoside,a4paper,cmspaper,final,collab]{cms-tdr}
\def\svnVersion{214205}\def\cmsCernNo{2013-037}\def\cmsCernDate{\today}\def\cmsMessage{Published in the European Physical Journal C as \href{http://dx.doi.org/10.1140/epjc/s10052-013-2568-6}{\doi{10.1140/epjc/s10052-013-2568-6.}}}

\begin{document}\cmsNoteHeader{SUS-12-028}

\hyphenation{had-ron-i-za-tion}
\hyphenation{cal-or-i-me-ter}
\hyphenation{de-vices}
\RCS$HeadURL: svn+ssh://alverson@svn.cern.ch/reps/tdr2/papers/SUS-12-028/trunk/SUS-12-028.tex $
\RCS$Id: SUS-12-028.tex 206958 2013-09-13 14:21:19Z bainbrid $

\newcommand\T{\rule{0pt}{2.6ex}}
\newcommand\B{\rule[-1.2ex]{0pt}{0pt}}
\newcommand{\mparent}{\ensuremath{m_{\text{parent}}}\xspace}
\newcommand{\msqgl}{\ensuremath{m_{\sQua/\sGlu}}\xspace}
\newcommand{\msq}{\ensuremath{m_{\sQua}}\xspace}
\newcommand{\mgl}{\ensuremath{m_{\sGlu}}\xspace}
\newcommand{\mstop}{\ensuremath{m_{\sTop}}\xspace}
\newcommand{\msbot}{\ensuremath{m_{\sBot}}\xspace}
\newcommand{\mlsp}{\ensuremath{m_{\mathrm{LSP}}}\xspace}
\newcommand{\chizz}{\PSGczDo\xspace}
\newcommand{\ra}{\ensuremath{\rightarrow}}
\newcommand{\Ttwo}{\texttt{D1}\xspace}
\newcommand{\TtwoBB}{\texttt{D2}\xspace}
\newcommand{\TtwoTT}{\texttt{D3}\xspace}
\newcommand{\Tone}{\texttt{G1}\xspace}
\newcommand{\ToneBBBB}{\texttt{G2}\xspace}
\newcommand{\ToneTTTT}{\texttt{G3}\xspace}
\newcommand{\nb}{\ensuremath{n_{\mathrm{b}}}\xspace}
\newcommand{\nbreco}{\ensuremath{n_{\mathrm{b}}^{\text{reco}}}\xspace}
\newcommand{\njet}{\ensuremath{n_{\text{jet}}}\xspace}
\newcommand{\njetlow}{\ensuremath{2 \leq \njet \leq 3}\xspace}
\newcommand{\njethigh}{\ensuremath{\njet \geq 4}\xspace}
\newcommand{\rat}{\ensuremath{R_{\alpha_{\mathrm{T}}}}}
\newcommand{\RaT}{\rat\xspace}
\newcommand{\sq}{\PSQ}
\providecommand{\PSQt}{\ensuremath{\widetilde{\cmsSymbolFace{t}}}\xspace} 
\newcommand{\gl}{\PSg\xspace}
\newcommand{\scalht}{\mbox{$H_\text{T}$}\xspace}
\newcommand{\pfmet}{\mbox{$\eslash_\text{T}^{\text{PF}}$}\xspace}
\newcommand{\cls}{\mbox{CL$_\mathrm{S}$}\xspace}
\newcommand{\znunu}{\ensuremath{\cPZ \rightarrow \cPgn\cPagn}}
\newcommand{\wtaunu}{\ensuremath{\PW \rightarrow \Pgt\cPgn}}
\providecommand{\Et}{\ET}
\providecommand{\met}{\ETslash}
\newcommand{\Hslash}{{\hbox{$H$\kern-0.8em\lower-.05ex\hbox{/}\kern0.10em}}}
\newcommand{\mht}{\mbox{$\Hslash_\mathrm{T}$}\xspace}
\newcommand{\dht}{\ensuremath{\Delta\scalht}\xspace}
\newcommand{\alphat}{\ensuremath{\alpha_{\text{T}}}\xspace}
\newcommand{\htalphat}{\texttt{HT\_AlphaT}\xspace}
\newcommand{\photon}{\texttt{Photon}\xspace}
\newcommand{\muht}{\texttt{Mu\_HT}\xspace}
\newcommand{\httrigger}{\texttt{HT}\xspace}
\newcommand{\htmht}{\texttt{HT\_MHT}\xspace}
\newcommand{\gj}{\ensuremath{\gamma + \text{jets}}\xspace}
\newcommand{\mj}{\ensuremath{\mu + \text{jets}}\xspace}
\newcommand{\mmj}{\ensuremath{\mu\mu + \text{jets}}\xspace}
\newcommand{\npre}{\ensuremath{N_{\text{pred}}}\xspace}
\newcommand{\nobs}{\ensuremath{N_{\text{obs}}}\xspace}
\newcommand{\njets}{\ensuremath{N_{\tex{jet}}}\xspace}
\providecommand{\re}{\ensuremath{\cmsSymbolFace{e}}}

\newcommand{\sQuaL}{\ensuremath{\widetilde{\cmsSymbolFace{q}}_\cmsSymbolFace{L}}\xspace}
\newcommand{\sQuaR}{\ensuremath{\widetilde{\cmsSymbolFace{q}}_\cmsSymbolFace{R}}\xspace}
\newcommand{\sU}{\ensuremath{\widetilde{\cmsSymbolFace{u}}\xspace}}
\newcommand{\sD}{\ensuremath{\widetilde{\cmsSymbolFace{d}}\xspace}}
\newcommand{\sC}{\ensuremath{\widetilde{\cmsSymbolFace{c}}\xspace}}
\newcommand{\sS}{\ensuremath{\widetilde{\cmsSymbolFace{s}}\xspace}}

\newlength\cmsFigWidth
\ifthenelse{\boolean{cms@external}}{\setlength\cmsFigWidth{0.85\columnwidth}}{\setlength\cmsFigWidth{0.4\textwidth}}
\ifthenelse{\boolean{cms@external}}{\providecommand{\cmsLeft}{top}}{\providecommand{\cmsLeft}{left}}
\ifthenelse{\boolean{cms@external}}{\providecommand{\cmsRight}{bottom}}{\providecommand{\cmsRight}{right}}

\cmsNoteHeader{SUS-12-028}

\title{Search for supersymmetry in hadronic final states with missing
transverse energy using the variables \alphat and b-quark multiplicity
in pp collisions at $\sqrt{s} = 8\TeV$}
\titlerunning{Supersymmetry with \ETslash using \alphat and b-quark multiplicity} \date{\today}

\abstract{An inclusive search for supersymmetric processes that
  produce final states with jets and missing transverse energy is
  performed in pp collisions at a centre-of-mass energy of 8\TeV. The
  data sample corresponds to an integrated luminosity of 11.7\fbinv
  collected by the CMS experiment at the LHC. In this search, a
  dimensionless kinematic variable, \alphat, is used to discriminate
  between events with genuine and misreconstructed missing transverse
  energy. The search is based on an examination of the number of
  reconstructed jets per event, the scalar sum of transverse energies
  of these jets, and the number of these jets identified as
  originating from bottom quarks. No significant excess of events over
  the standard model expectation is found. Exclusion limits are set in
  the parameter space of simplified models, with a special emphasis on
  both compressed-spectrum scenarios and direct or gluino-induced
  production of third-generation squarks. For the case of
  gluino-mediated squark production, gluino masses up to 950--1125\GeV
  are excluded depending on the assumed model. For the direct
  pair-production of squarks, masses up to 450\GeV are excluded for a
  single light first- or second-generation squark, increasing to
  600\GeV for bottom squarks.
}

\hypersetup{
pdfauthor={CMS Collaboration},
pdftitle={Search for supersymmetry in hadronic final states with missing transverse energy using the variables AlphaT and b-quark multiplicity in pp collisions at 8 TeV},
pdfsubject={CMS, physics, jets, missing energy, supersymmetry, SUSY,
AlphaT},
pdfkeywords={CMS, physics, jets, missing energy, supersymmetry, SUSY,
AlphaT}, }

\maketitle

\section{Introduction}

The standard model (SM) of particle physics has been extremely
successful in describing phenomena at the highest energies attained
thus far. Nevertheless, it is widely believed to be only an effective
approximation of a more complete theory that would supersede the SM at a
higher energy scale. Supersymmetry (SUSY)~\cite{ref:SUSY-1, ref:SUSY0,
  ref:SUSY1, ref:SUSY2, ref:SUSY3, ref:SUSY4, ref:hierarchy1,
  ref:hierarchy2} is generally regarded as one of the likely
extensions to the SM. The theory is based on the unique way to extend
the space-time symmetry group underpinning the SM, introducing a
relationship between fermions and bosons.

A low-energy realisation of SUSY, \eg at the TeV scale, is motivated
by the cancellation of the quadratically divergent loop corrections to
the Higgs boson mass in the SM~\cite{ref:hierarchy1, ref:hierarchy2}.
In order to avoid a large amount of fine-tuning in these loop
corrections, the difference in masses between the top quark and the
third-generation squarks must not be too
large~\cite{ref:barbierinsusy}. While the majority of SUSY particles
(sparticles) may be beyond the reach of the Large Hadron Collider
(LHC) at the present beam energy and luminosity, the recent discovery
of a low-mass Higgs boson candidate~\cite{ref:atlashiggsdiscovery,
  ref:cmshiggsdiscovery} motivates ``natural'' SUSY models in which
top and bottom squarks (and gluinos) appear at the TeV scale. For
R-parity-conserving SUSY~\cite{Farrar:1978xj}, sparticles will be
produced in pairs and decay to SM particles and the lightest sparticle
(LSP), which is generally assumed to be weakly interacting and
massive. Therefore, the pair production of massive coloured sparticles
is expected to result in a signature that is rich in jets, in
particular those originating from bottom quarks if the
third-generation squarks are light, and contains a significant amount
of missing transverse energy, \met, due to the undetected LSPs.

This paper summarises an inclusive search for pair production of
massive coloured sparticles in final states with jets and \met,
performed in pp collisions at a centre-of-mass energy $\sqrt{s} =
8\TeV$. The analysed data sample corresponds to an integrated
luminosity of $11.7 \pm 0.5\fbinv$~\cite{lumi} collected by the
Compact Muon Solenoid (CMS) experiment. Several other searches in this
channel have been conducted by both the ATLAS and CMS
experiments~\cite{atlas-0, atlas-1, atlas-2, atlas-3, atlas-4, atlas-5, cms-1,
  cms-2, cms-3, cms-4, RA1Paper2011FULL, RA1Paper2011, RA1Paper}. The
strategy of the analysis presented in this paper is based on the
kinematic variable \alphat, which provides powerful discrimination
against multijet production, a manifestation of quantum chromodynamics
(QCD), while maintaining sensitivity to a wide range of SUSY
models. This analysis extends previous searches based on a similar
strategy with samples of pp collisions at $\sqrt{s} =
7\TeV$~\cite{RA1Paper2011FULL, RA1Paper2011, RA1Paper}.

In order to improve the sensitivity of the analysis to the main
production mechanisms of massive coloured sparticles at hadron
colliders (squark-squark, squark-gluino, and gluino-gluino), events
with significant \met and two or more energetic jets are categorised
according to the jet multiplicity. Events with two or three
reconstructed jets are used to search for squark-squark and
squark-gluino production, while events with four or more reconstructed
jets probe gluino-gluino production. This classification according to
the jet multiplicity is a new feature with respect to the previous
analysis~\cite{RA1Paper2011FULL}. Moreover, to enhance the sensitivity
to third-generation squark signatures, events are further categorised
according to the number of reconstructed jets identified as
originating from bottom quarks (b-quark jets). The analysis also
considers a large dynamic range in the scalar sum of the transverse
energies of reconstructed jets in order to probe signal models over a
large range of mass splittings between the parent sparticle and the
LSP, including models characterised by a compressed
spectrum~\cite{compressed-spectra}. This approach provides sensitivity
to a wide variety of SUSY event topologies arising from the pair
production and decay of massive coloured sparticles while still
maintaining the character of an inclusive search.

\section{Interpretation with simplified models}

To interpret the results of this search, simplified
models~\cite{Alwall:2008ag,Alwall:2008va,sms} are used. These
effective models include only a limited set of sparticles (production
and decay) to enable comprehensive studies of individual SUSY event
topologies.
The result of this search can also be interpreted in a range of other
relevant models, such as the constrained minimal supersymmetric
extension of the standard model (CMSSM)~\cite{ref:MSUGRA,
  PhysRevLett.69.725, ref:CMSSM} or other effective or complete SUSY
models that predict event topologies with two or more energetic jets
and significant \met.

In this paper, we focus on the interpretation in two classes of
simplified models, the first of which describes direct pair production
of squarks, including top and bottom squarks, that decay to a quark of
the same flavour and the LSP. The second class describes
gluino-induced production of (off-shell) squarks, again including top
and bottom squarks, in which gluino pair production is followed by the
decay of each gluino to a quark-antiquark pair and the LSP. The
simplified models considered in this analysis are summarised in
Table~\ref{tab:simplified-models}. For each model, the LSP is assumed
to be the lightest neutralino.

Table~\ref{tab:simplified-models} also defines reference models in
terms of the parent (gluino or squark) and LSP sparticle masses,
$m_\text{parent}$ and $m_\mathrm{LSP}$, respectively, which are used to
illustrate potential yields in the signal region. In the case of the
model \TtwoTT, a massless LSP is considered. The masses are chosen to
be reasonably high while still being within the expected sensitivity
reach.

\begin{table}[htb]
  \topcaption{A summary of the simplified models considered in this
    analysis, which involve both direct (\texttt{D}) and
    gluino-induced (\texttt{G}) production of squarks, and their
    decays. Models \Ttwo and \Tone concern the direct or
    gluino-induced production of first- or second-generation squarks
    only. Reference models are also defined in terms of the parent
    (gluino or squark) and LSP sparticle masses. }
  \label{tab:simplified-models}
  \setlength{\extrarowheight}{2.5pt}
  \centering
  \begin{tabular}{ llccc }
    \hline
    Model & Production/decay mode & \multicolumn{2}{c}{Reference model} \\
          &                       & \mparent  & \mlsp    \\
          &                       & [\GeVns{}]                    & [\GeVns{}]                   \\ [0.5ex]
    \hline
    \Ttwo     & $\Pp\Pp\,\ra\,\sQua\sQua^{*}\,\ra\,\cPq\chizz\,\cPaq\chizz$                   & 600 & 250 \\
    \TtwoBB   & $\Pp\Pp\,\ra\,\sBot\sBot^{*}\,\ra\,\cPqb\chizz\,\cPaqb\chizz$                   & 500 & 150 \\
    \TtwoTT   & $\Pp\Pp\,\ra\,\sTop\sTop^{*}\,\ra\,\cPqt\chizz\,\cPaqt\chizz$                   & 400 & 0   \\
    \Tone     & $\Pp\Pp\,\ra\,\sGlu\sGlu\,\ra\,\cPq\cPaq\chizz\cPq\,\cPaq\chizz$ & 700 & 300 \\
    \ToneBBBB & $\Pp\Pp\,\ra\,\sGlu\sGlu\,\ra\,\cPqb\cPaqb\chizz\,\cPqb\cPaqb\chizz$ & 900 & 500 \\
    \ToneTTTT & $\Pp\Pp\,\ra\,\sGlu\sGlu\,\ra\,\cPqt\cPaqt\chizz\,\cPqt\cPaqt\chizz$ & 850 & 250 \\
    \hline
  \end{tabular}
\end{table}

\section{The CMS apparatus}

{\tolerance=600 The central feature of the CMS apparatus is a superconducting solenoid
of 6\unit{m} internal diameter, providing a magnetic field of
3.8\unit{T}. Within the superconducting solenoid volume are a silicon
pixel and strip tracker, an electromagnetic calorimeter (ECAL)
comprising 75\,848 lead-tungstate crystals, and a brass/scintillator
hadron calorimeter (HCAL). Muons are measured in gas-ionisation
detectors embedded in the steel flux return yoke of the
magnet. Extensive forward calorimetry complements the coverage
provided by the barrel and endcap detectors.  The CMS detector is
nearly hermetic, which allows for momentum balance measurements in the
plane transverse to the beam axis.\par}

CMS uses a right-handed coordinate system, with the origin at the
nominal interaction point, the $x$ axis pointing to the centre of the
LHC ring, the $y$ axis pointing up (perpendicular to the plane of the
LHC ring), and the $z$ axis along the anticlockwise-beam
direction. The polar angle $\theta$ (radians) is measured from the
positive $z$ axis and the azimuthal angle $\phi$ (radians) is measured
in the $x$-$y$ plane. Pseudorapidity is defined as $\eta = -\ln [ \tan
(\theta/2)]$.

The silicon pixel and strip tracking systems measure charged particle
trajectories with full azimuthal coverage and a pseudorapidity
acceptance of $|\eta| < 2.5$. The resolutions on the transverse
momentum (\pt) and impact parameter of a charged particle with $\pt <
40\gev$ are typically 1\% and 15\mum, respectively. Muons are
measured in the pseudorapidity range $\abs{\eta}< 2.4$. Matching muons
to tracks measured in the tracking subdetectors results in a
\pt resolution between 1 and 5\% for $\pt \leq 1\TeV$.

The ECAL has an energy resolution of better than 0.5\% for unconverted
photons with transverse energies above 100\GeV. The HCAL, when
combined with the ECAL, measures jets with a resolution $\Delta E/E
\approx 100\% / \sqrt{E\,[\GeVns]} \oplus 5\%$.  In the region $\abs{
  \eta }< 1.74$, the HCAL cells have widths of 0.087 in pseudorapidity
and 0.087 in azimuth. In the $\eta$-$\phi$ plane, and for $\abs{\eta}<
1.48$, the HCAL cells map onto $5 \times 5$ arrays of ECAL crystals to
form calorimeter towers projecting radially outwards from close to the
nominal interaction point. At larger values of $\abs{ \eta }$, the
size of the towers increases and the matching ECAL arrays contain
fewer crystals. Within each tower, the energy deposits in ECAL and
HCAL cells are summed to define the calorimeter tower energies,
subsequently used to provide the energies and directions of hadronic
jets.

The first level (L1) of the CMS trigger system, composed of custom
hardware processors, uses information from the calorimeters and muon
detectors to select the most interesting events in a fixed time
interval of less than 4\mus. The high-level trigger (HLT) processor
farm further decreases the event rate, from around 100\unit{kHz} to
around 300\unit{Hz}, before data are stored.

A more detailed description of the CMS detector can be found in
Ref.~\cite{ref:CMS}.
\section{Event reconstruction and selection\label{sec:event}}

\subsection{Definition of \texorpdfstring{\alphat}{AlphaT}\label{sec:alphat}}

The \alphat~\cite{Randall:2008rw, RA1Paper} variable is used to reject
multijet events efficiently without significant \met or with
transverse energy mismeasurements, while retaining a large sensitivity
to new physics with final-state signatures containing significant
\met.

The measurement of \met typically relies on independent sources of
information from each of the calorimeter, tracking, and muon
subdetectors~\cite{cms-met}. Relative to other physics objects, this
measurement is particularly sensitive to the beam conditions and
detector performance. This difficulty is compounded by the
high-energy, high-luminosity hadron collider environment at the LHC
and the lack of precise theoretical predictions for the kinematic
properties and cross sections of multijet events.

Given these difficulties, the variable \alphat was
developed to avoid direct reliance on a measurement of \met, instead
depending solely on the measurements of the transverse energies and
(relative) azimuthal angles of jets, which are reconstructed from
energy deposits in the calorimeter towers~\cite{Chatrchyan:2011ds}.
The variable is intrinsically robust against the presence of jet
energy mismeasurements in multijet systems.
For dijet events, the \alphat variable is defined
as~\cite{Randall:2008rw, RA1Paper}:
\begin{equation}
\label{eq:alphat}
\alphat =\frac{\Et^{\mathrm{j}_2}}{M_\mathrm{T}}\, ,
\end{equation}
where $\Et^{\mathrm{j}_2}$ is the transverse energy of the less energetic
jet and $M_\text{T}$ is the transverse mass of the dijet system,
defined as
\begin{equation}
  \label{eq:mt}
  M_\text{T} = \sqrt{ \left( \sum_{i=1}^2 \Et^{\mathrm{j}_i}
    \right)^2 - \left( \sum_{i=1}^2 p_x^{\mathrm{j}_i} \right)^2 - \left(
      \sum_{i=1}^2 p_y^{\mathrm{j}_i} \right)^2}\, .
\end{equation}
where $\Et^{\mathrm{j}_i}$, $p_x^{\mathrm{j}_i}$, and $p_y^{\mathrm{j}_i}$ are,
respectively, the transverse energy and $x$ or $y$ components of the
transverse momentum of jet $\mathrm{j}_i$.

For a perfectly measured dijet event with $\Et^{\mathrm{j}_1} =
\Et^{\mathrm{j}_2}$ and jets back-to-back in $\phi$, and in the limit
in which the momentum of each jet is large compared with its mass, the
value of \alphat is 0.5. For the case of an imbalance in the measured
transverse energies of back-to-back jets, \alphat is reduced to a
value smaller than 0.5, which gives the variable its intrinsic
robustness with respect to jet energy mismeasurements.  A similar
behaviour is observed for energetic dijet events that contain
neutrinos from the decay of a bottom or charm quark, as the neutrinos
are typically collinear with respect to the axis of the heavy-flavour
jet. Values significantly greater than 0.5 are observed when the two
jets are not back-to-back and are recoiling against significant,
genuine \met.

The definition of the \alphat variable can be generalised for events
with two or more jets as follows. The mass scale of the physics
processes being probed is characterised by the scalar sum of the
transverse energy $\Et$ of jets considered in the analysis, defined as
$\scalht = \sum_{i=1}^{\njet} \Et^{\mathrm{j}_i}$, where \njet is the
number of jets with \Et above a predefined threshold. The estimator
for \met is given by the magnitude of the transverse momenta
$\vec{\pt}$ vectorial sum over these jets, defined as $\mht =
|\sum_{i=1}^{N_\text{jet}} \vec{\pt}^{\mathrm{j}_i}|$.
For events with three or more jets, a pseudo-dijet system is formed by
combining the jets in the event into two pseudo-jets. The total \Et
for each of the two pseudo-jets is calculated as the scalar sum of the
measured \Et of the contributing jets. The combination chosen is the
one that minimises the absolute \Et difference between the two
pseudo-jets, \dht. This simple clustering criterion provides the best
separation between multijet events and events with genuine
\met. Equation~(\ref{eq:alphat}) can therefore be generalised as:
\begin{equation}
  \label{eq:alphat2}
  \alphat = \frac{1}{2} \times \frac{\scalht -
    \dht}{\sqrt{\scalht^2 - \mht^2}}  = \frac{1}{2} \times
  \frac{1 - (\dht/\scalht)}{\sqrt{1 - (\mht/\scalht)^2}}\,.
\end{equation}

In the presence of jet energy mismeasurements or neutrinos from
heavy-flavour quark decays, the direction and magnitude of the
apparent missing transverse energy, \mht, and energy imbalance of the
pseudo-dijet system, \dht, are highly correlated. This correlation is
much weaker for R-parity-conserving SUSY with each of the two decay
chains producing the LSP.

\subsection{Physics objects \label{sec:objects}}

Jets are reconstructed from the energy deposits in the calorimeter
towers~\cite{Chatrchyan:2011ds}, clustered by the infrared-safe
anti-\kt algorithm~\cite{antikt} with a size parameter of
0.5. In this process, the contribution from each calorimeter tower is
assigned a momentum, the absolute value and the direction of which are
given by the energy measured in the tower and the position of the
tower. The raw jet energy is obtained from the sum of the tower
energies and the raw jet momentum by the vectorial sum of the tower
momenta, which results in a nonzero jet mass. The raw jet energies are
corrected to remove the effects of overlapping pp collisions
(pileup)~\cite{Cacciari2008119, 1126-6708-2008-04-005} and to
establish a relative uniform response of the calorimeter in $\eta$ and
a calibrated absolute response in \pt.

The presence of a b-quark jet is inferred by the Combined Secondary
Vertex algorithm~\cite{CMS-PAS-BTV-12-001} that incorporates several
measurements to build a single discriminating variable that can be
used to identify jets originating from bottom quarks with high
efficiency and purity.
Due to the pixel-detector acceptance, b-quark jets are identified only
in the region $|\eta| < 2.4$. In this analysis, the discriminator
threshold is chosen such that the probability to misidentify (mistag)
jets originating from light-flavour partons (u, d, s quarks or gluons)
as b-quark jets is approximately 1\% for jets with transverse momenta
of 80\GeV~\cite{CMS-PAS-BTV-12-001}. This threshold results in a
b-tagging efficiency, \ie the probability to correctly identify jets
as originating from bottom quarks, in the range
60--70\%~\cite{CMS-PAS-BTV-12-001}, dependent on jet \pt.

The reconstruction of photons, electrons and muons is described
below. The presence (or absence) of these objects is used to define
the event samples for the signal and multiple control regions, the
latter of which are used to estimate the background contributions from
SM processes in the signal region.

The energy of photons~\cite{PAS-EGM-10-006} is directly obtained from
the ECAL measurement, corrected for zero-suppression and pileup
effects. Various identification criteria must be met in order to
correctly identify photons with high efficiency and suppress the
misidentification of electrons, jets, or spurious ECAL noise as
photons. These include the requirements that the shower shape of the
energy deposition in the ECAL be consistent with that expected from a
photon, the energy detected in the HCAL behind the photon shower must
not exceed 5\% of the photon energy, and no matched hits in the pixel
tracker must be found. Isolation from other activity in the event is
determined through a combination of independent energy sums obtained
from each of the HCAL, ECAL, and tracker subdetectors within a cone of
$\Delta R = \sqrt{(\Delta\phi)^2 + (\Delta\eta)^2} = 0.3$ around the
photon trajectory. These sums are corrected for pileup effects and for
the contributions from the photon itself.

The energy of electrons~\cite{PAS-EGM-10-004} is determined from a
combination of the track momentum at the main interaction vertex, the
corresponding ECAL cluster energy, and the energy sum of all
bremsstrahlung photons attached to the track. Identification criteria
similar to those described above for photons are applied, with
additional requirements on the associated track that consider the
track quality, energy-momentum matching, and compatibility with the
main interaction vertex in terms of the transverse and longitudinal
impact parameters.

The energy of muons~\cite{PAS-MUO-10-004} is obtained from the
corresponding track momentum, combining measurements from the muon
detectors and both the silicon pixel and strip tracking
subdetectors. Various track quality criteria are considered when
identifying muons, as are the transverse and longitudinal impact
parameters with respect to the main interaction vertex.

Isolation of muons and electrons is based on a combination of
independent sums from the HCAL, ECAL, and tracker subdetectors and
measured relative to the muon or electron transverse momentum. The
isolation sums are determined for a cone of radius $\Delta R = 0.3$
(0.4) around the electron (muon) trajectory and are corrected for the
effects of pileup and for the contributions from the lepton itself.

\subsection{Event selection for the signal region\label{sec:signal}}

Events containing non-collision backgrounds are suppressed by
requiring at least one vertex of high-\pt tracks to be reconstructed
in the luminous region.
In the case of multiple vertices, the main interaction vertex is
defined as the one with the highest scalar sum of ${\pt}^{2}$ of all
associated tracks.

In order to suppress SM processes with genuine \met from neutrinos in
the final state, events are vetoed if they contain an isolated
electron or muon with $\pt > 10\GeV$. Events with an isolated photon
with $\pt > 25\GeV$ are also vetoed to ensure an all-jet final state.

Jets are required to have transverse energy $\Et > 50\GeV$ and $|\eta|
< 3.0$. The two highest-$\Et$ jets must each satisfy $\Et > 100\GeV$.
These two \Et\ requirements are relaxed for some signal regions, as
described below. 
The highest-$\Et$ jet is additionally required to satisfy $|\eta| <
2.5$. Events are vetoed that contain rare, spurious signals from the
calorimeters~\cite{1748-0221-5-03-T03014} that are misidentified as
jets. To ensure that the variable \mht is an unbiased estimator of
\met, events are vetoed if any additional jet satisfies both $\Et >
50\GeV$ and $|\eta| > 3$.

Events are required to have $\scalht > 275\GeV$ to ensure high
efficiency for the trigger conditions used to record the event sample,
described in Section~\ref{sec:trigger}. The signal region is divided
into eight bins in \scalht: two bins of width $50\GeV$ in the range
$275 < \scalht < 375\GeV$, five bins of width $100\GeV$ in the range
$375 < \scalht < 875\GeV$, and a final open bin, $\scalht >
875\GeV$. As in Ref.~\cite{RA1Paper}, the jet $\Et$ threshold is
scaled down to 37 and 43\GeV for the regions $275 < \scalht < 325$
and $325 < \scalht < 375\GeV$, respectively. The threshold for the two
highest-\Et jets is also scaled accordingly to $73$ and $87\GeV$. This
is done in order to maintain a background composition similar to that
observed for the higher \scalht bins, and also to increase the
analysis acceptance for SUSY models characterised by compressed
spectra.

Events are further categorised according to the number of jets per
event, \njetlow or \njethigh, and the number of reconstructed b-quark
jets per event, $\nb = 0$, 1, 2, 3, or $\geq$4.  For the category of
events satisfying $\njet \geq 4$ and $\nb \geq 4$, the six highest
\scalht bins are combined to give a final open bin of $\scalht >
375\GeV$.

For events satisfying the selection criteria described above, the
multijet background dominates over all other SM backgrounds. As
discussed in Section~\ref{sec:alphat}, multijet events populate the
region $\alphat < 0.5$.
The \alphat distribution is characterised by a sharp edge at 0.5,
beyond which the multijet event yield falls by several orders of
magnitude. Multijet events with extremely rare but large stochastic
fluctuations in the calorimetric measurements of jet energies can lead
to values of \alphat slightly above 0.5. The edge at 0.5 sharpens with
increasing \scalht for multijet events, primarily due to a
corresponding increase in the average jet energy and thus an
improvement in the jet energy resolution. This effect yields an
exponential dependence on \scalht for the ratio of multijet events
with a value of \alphat above and below a given threshold value
(larger than 0.5), as described further in Section~\ref{sec:multijet}.

The contribution from multijet events is suppressed by many orders of
magnitude by requiring $\alphat > 0.55$. As an example, an event that
satisfies both $\scalht = 275\, (875)\GeV$ and $\alphat = 0.55$ must
also satisfy $\mht \geq 115\,$ (365)\GeV. However, certain classes of
rare background events can lead to values of \alphat greater than
0.55, such as those containing beam halo, reconstruction failures,
spurious detector noise, or event misreconstruction due to detector
inefficiencies. These event classes, with large, non-physical values
of \met, are rejected by applying dedicated vetoes~\cite{cms-met}, the
most important of which are described below.

The first example concerns events containing severe energy
mismeasurements as a result of jets being reconstructed within or near
to inefficient regions in the ECAL (which amount to $\sim$1\% of the
ECAL channel count) or the instrumentation gap between the ECAL barrel
and endcap systems at $|\eta| = 1.48$. These events are identified and
vetoed as follows. The negative vector sum of jet transverse momenta
when jet $j$ is ignored, defined as $-\sum_{i = 1, i \neq j}^{\njet}
\vec{\pt}^{i}$, is determined for each ignored jet in turn, $1 \leq j
< \njet$. An azimuthal distance of $\Delta\phi < 0.5$ between the
directions of jet $j$ and the corresponding vector sum indicates that
jet $j$ has suffered a sufficiently large energy mismeasurement to
satisfy $\alphat > 0.55$. The event is rejected if the angular
distance in the ($\eta,\phi$) plane between the affected jet and the
closest inefficient ECAL region satisfies $\Delta R < 0.3$. Similarly,
the event is rejected if the $\eta$ position of the affected jet
satisfies $\Delta\eta < 0.3$ with respect to the ECAL barrel-endcap
boundary.

The second example concerns the rare circumstance in which several
jets with transverse energies below the \Et thresholds and aligned in
$\phi$ result in significant \mht relative to the value of \met (which
is less sensitive to jet \Et thresholds). This type of background,
typical of multijet events, is suppressed while maintaining high
efficiency for SM or new physics processes with genuine, significant
\met by requiring $\mht / \met < 1.25$. The measurement of \met is
provided by the particle-flow (PF) reconstruction
framework~\cite{CMS-PAS-PFT-09-001, CMS-PAS-PFT-10-001}.

\begin{figure*}[!h]
  \begin{center}
    \includegraphics[width=0.48\textwidth]{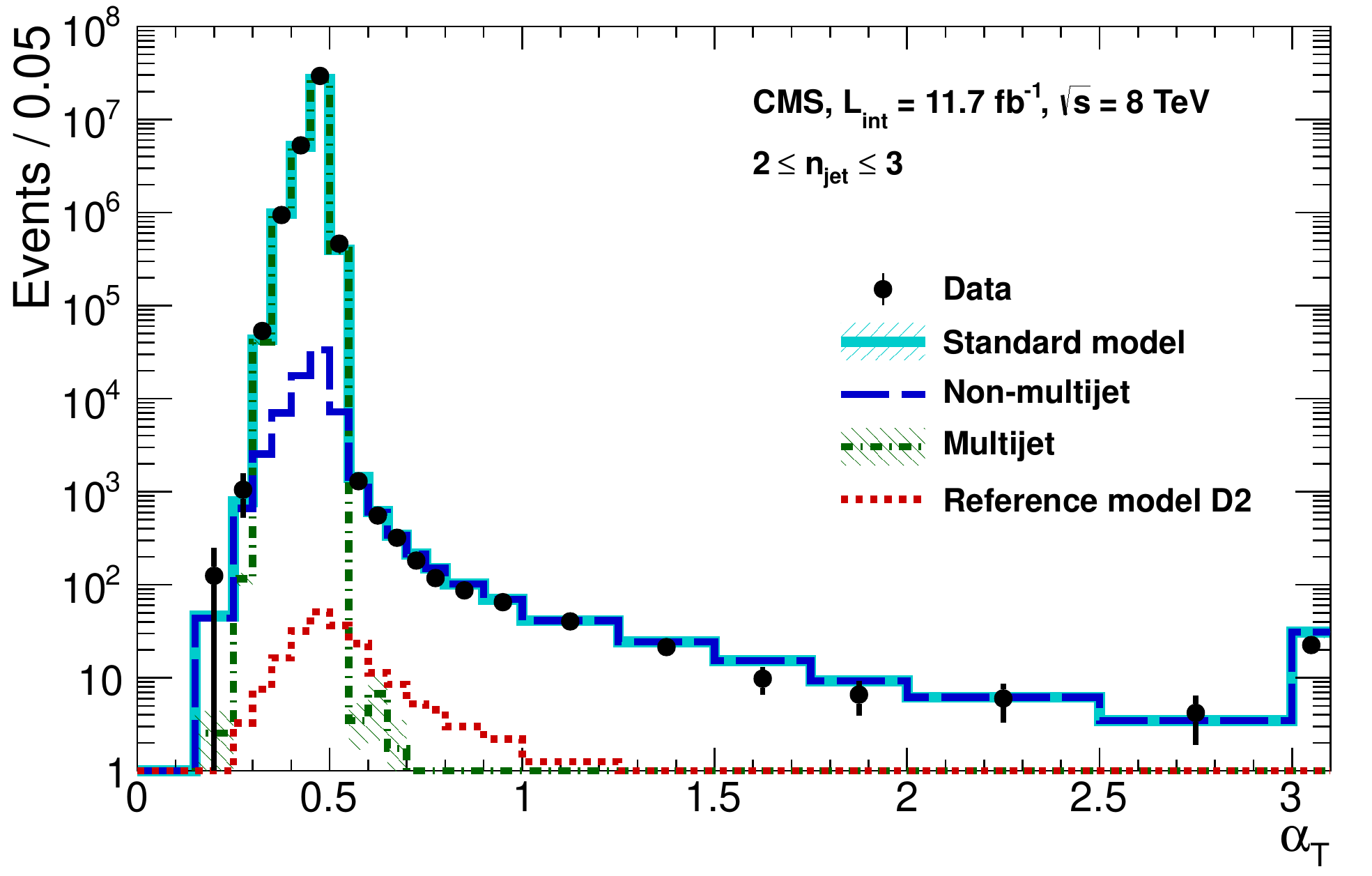}
    \includegraphics[width=0.48\textwidth]{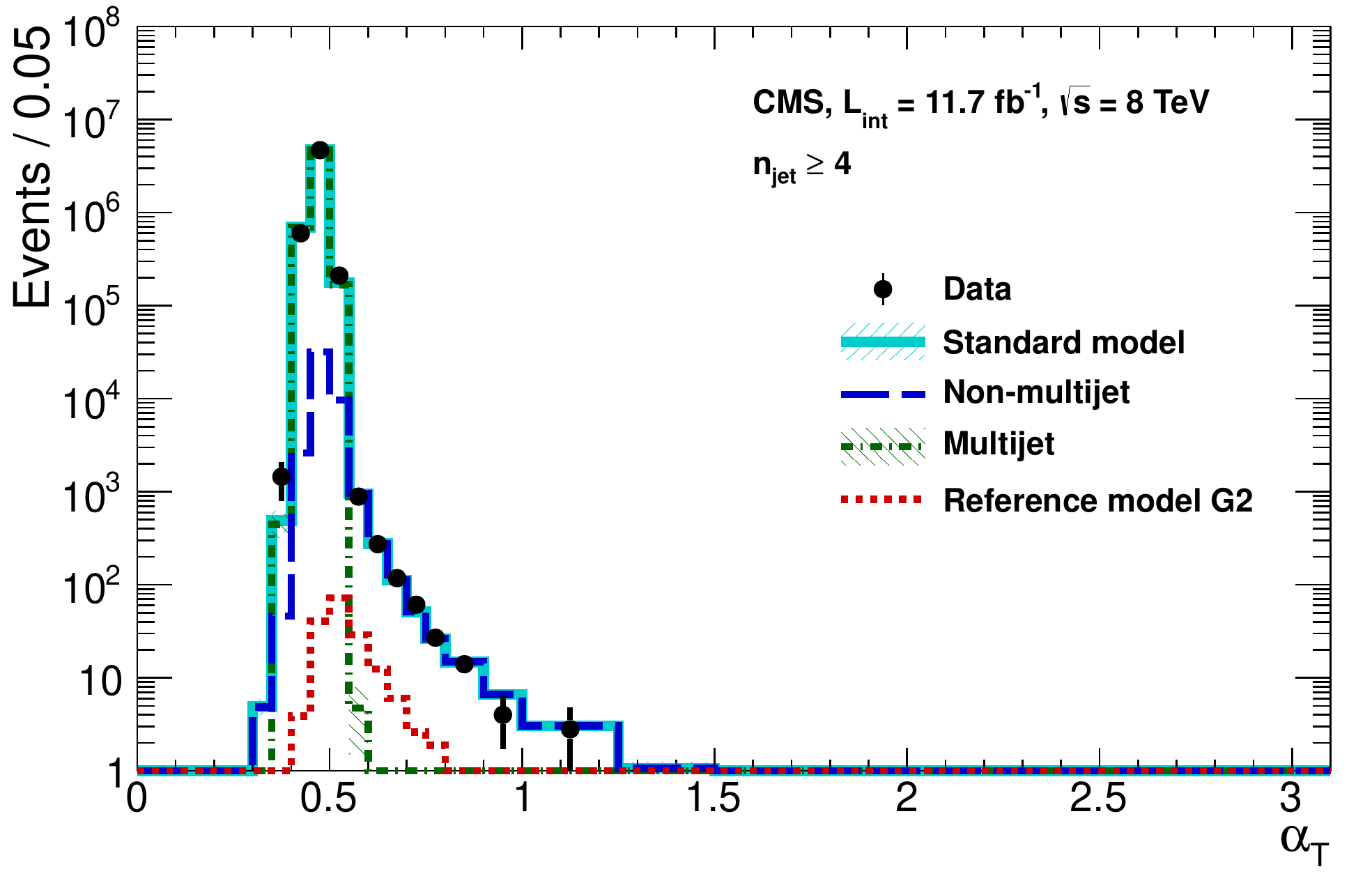} \\
    \caption{The \alphat distributions of events with $\scalht > 375\GeV$ that satisfy all the selection criteria described above
      except the \alphat requirement, categorised according to
      \njetlow (left) and \njethigh (right). An inclusive set of
      trigger conditions is used to collect the events in data (black
      solid circles with error bars). Expected yields as given by
      simulation are also shown for multijet events (green dash-dotted
      line), non-multijet backgrounds with genuine \met as described
      in Section~\ref{sec:non-multijet} (blue long-dashed line), the
      sum of all aforementioned SM processes (cyan solid line) and the
      reference signal model \TtwoBB (left, red dotted line) or
      \ToneBBBB (right, red dotted line). The statistical
      uncertainties for the multijet and SM expectations are
      represented by the hatched areas (visible only for statistically
      limited bins). The final bin contains all events with $\alphat >
      3$. (Colour figure online.) }
    \label{fig:alphaT}
    \end{center}
\end{figure*}

Figure~\ref{fig:alphaT} shows the \alphat distributions of events with
$\scalht > 375\GeV$ that satisfy all the selection criteria described
above except the \alphat requirement, categorized according to
\njet. An inclusive set of trigger conditions is used in order to show
the full \alphat distribution.
The analysis relies on data control samples to estimate the
contributions from the multijet and non-multijet backgrounds, as
described in Sections~\ref{sec:non-multijet} and
\ref{sec:multijet}. However, for illustration, the expected yields
from simulation of multijet events, non-multijet backgrounds with
genuine \met, the sum of these SM backgrounds, and an example signal
model, are also shown in Fig.~\ref{fig:alphaT}. The expected yield for
multijet events that satisfy $\alphat > 0.55$, as given by simulation,
is less than ten events and is negligible with respect to all other SM
backgrounds. Figure~\ref{fig:alphaT} highlights the ability of the
\alphat variable to discriminate between multijet events and all other
SM or new physics processes with genuine \met in the final state.

\subsection{Trigger conditions\label{sec:trigger}}

Events are recorded with multiple jet-based trigger conditions,
implemented on the HLT computing farm, that require both \scalht and
\alphat to lie above predetermined thresholds, as summarised in
Table~\ref{tab:triggers}. Different trigger conditions are used
depending on the analysis \scalht bin.  The trigger-level jet energies
are corrected to account for scale and pileup effects.
The thresholds used in the \scalht binning scheme are shifted up by
25\GeV with respect to the trigger thresholds in order to maintain
high efficiency for the \scalht component of the trigger condition.

The trigger efficiency, defined as the probability with which events
that satisfy the signal region selection criteria also satisfy the
trigger condition, is measured from data for each \njet category.  The
efficiency is measured using a data sample of $\mu + \text{jets}$
events recorded by an independent and unbiased trigger condition that
requires an isolated muon satisfying $\pt > 24\GeV$ and
$|\eta|<2.1$. The muon is required to be well separated from the
nearest jet $j$ by requiring $\Delta R(\mu,j) > 0.5$ and is ignored in
the calculation of \scalht and \alphat in order to emulate a genuine
\met signature.

The measured efficiencies are summarised in Table~\ref{tab:triggers}.
Non-negligible inefficiencies, which are accounted for in the final
result, are observed only for the lowest \scalht bin. The HLT-based
trigger conditions are dependent on multiple requirements on
quantities determined by the L1 trigger logic, which include
combinations of scalar sums of jet \Et measurements and individual \Et
thresholds on sub-leading jets. The different efficiencies measured
for the two \njet categories in the lowest \scalht bin are a result of
the requirements on L1 trigger quantities that exhibit non-negligible
inefficiencies at very low \scalht.

\begin{table}[htb]
  \topcaption{Trigger conditions used to record events for each \scalht
    bin and their efficiencies (with statistical uncertainties)
    measured in data for each \scalht bin and \njet category. }
  \label{tab:triggers}
  \centering
  \begin{tabular}{ ccccc }
    \hline
    Analysis bin & \multicolumn{2}{c}{Trigger thresholds} & \multicolumn{2}{c}{Trigger efficiency [\%]} \\
    \scalht [\GeVns{}] & \scalht [\GeVns{}] & \alphat & $2 \leq \njet \leq 3$           & $\njet \geq 4$                  \\
    \hline
    275--325\T    & 250           & 0.55    & $\phantom{1}89.1^{+0.4}_{-0.4}$ & $\phantom{1}83.7^{+0.6}_{-0.6}$ \\
    325--375\T    & 300           & 0.53    & $\phantom{1}98.7^{+0.2}_{-0.3}$ & $\phantom{1}98.2^{+0.4}_{-0.5}$ \\
    375--475\T    & 350           & 0.52    & $\phantom{1}99.0^{+0.4}_{-0.5}$ & $\phantom{1}99.7^{+0.2}_{-0.6}$ \\
    $\geq$475\T   & 400           & 0.51    & $100.0^{+0.0}_{-0.6}$           & $100.0^{+0.0}_{-0.8}$           \\
    \hline
  \end{tabular}
\end{table}

\section{Estimating the non-multijet backgrounds\label{sec:non-multijet}}

\subsection{Dominant background processes}

In the absence of a significant contribution from multijet events, the
remaining backgrounds in the signal region stem from SM processes with
significant \met in the final state.

For events in which no b-quark jets are identified, the largest
backgrounds are from the production of W and Z bosons in association
with jets. The decay \znunu\ is the only relevant contribution
from $\cPZ+\text{jets}$ events. For $\PW+\text{jets}$ events, the two relevant sources
are leptonic decays, in which the lepton is not reconstructed or fails
the isolation or acceptance requirements, and the decay $\wtaunu$
in which the $\tau$ decays hadronically and is identified as a jet.

For events satisfying $\nb \geq 1$, \ttbar production followed by
semileptonic decays becomes the most important background process. For
the subset of events satisfying $\nb = 1$ and \njetlow, the total
contribution from the $\PW+\text{jets}$ and $\cPZ+\text{jets}$ backgrounds is comparable
to the \ttbar background; otherwise \ttbar production
dominates. Events with three or more reconstructed b-quark jets
originate almost exclusively from \ttbar events, in which one or
several jets are misidentified as b-quark jets.

Residual contributions from single-top-quark and diboson production
are also expected.

\subsection{Definition of the data control samples}

Three independent data control samples, binned identically to the
signal region, are used to estimate the contributions from the various
background processes. These samples are defined by a selection of \mj,
\mmj, and \gj events. The event selection criteria for these control
samples are defined to ensure that any potential contamination from
multijet events is negligible. Furthermore, the selections are also
expected to suppress contributions from a wide variety of SUSY models
(signal contamination) to a negligible level. The selection criteria
that define the three data control samples are chosen such that the
composition of background processes and their kinematic properties
resemble as closely as possible those of the signal region once the
muon, dimuon system, or photon are ignored when computing quantities
such as \scalht, \dht, \mht, and \alphat. This approach emulates the
effects, including misreconstruction and acceptance, that lead to the
presence of these background processes in the signal region.

The $\mu + \text{jets}$ sample is recorded using a trigger condition
that requires an isolated muon satisfying $\pt > 24\GeV$ and
$|\eta|<2.1$.  The event selection requires exactly one isolated muon
that satisfies stringent quality criteria, $\pt > 30\GeV$, and
$|\eta|< 2.1$ in order for the trigger to be maximally efficient at
$(88.0 \pm 2.0)\%$.  Furthermore, the transverse mass of the muon and
\met~\cite{CMS-PAS-PFT-09-001, CMS-PAS-PFT-10-001} system must be
larger than $30\GeV$ to ensure a sample rich in W bosons. The muon is
required to be separated from the closest jet in the event by the
distance $\Delta R > 0.5$. The event is rejected if two muon
candidates are identified that have an invariant mass within a window
of $\pm 25\GeV$ around the mass of the Z boson, regardless of the
quality and isolation of the second muon candidate. No selection
requirement on \alphat is made in order to increase the statistical
precision of the predictions derived from this sample, while the
impact of removing the \alphat requirement is tested with a dedicated
set of closure tests described in Section~\ref{sec:syst}.

The $\mu\mu + \text{jets}$ sample uses the same trigger condition as
the $\mu + \text{jets}$ sample. Events are selected by requiring
exactly two oppositely charged, isolated muons that satisfy stringent
quality criteria and $|\eta|< 2.1$. The highest-\pt and lowest-\pt
muons must satisfy $\pt > 30\GeV$ and $\pt > 10\GeV$,
respectively. The invariant mass of the di-muon system is required to
be within a window of $\pm 25\GeV$ around the mass of the Z
boson. Both muons are required to be separated from their closest jets
in the event by the distance $\Delta R > 0.5$. Again, no requirement
on \alphat is made. These selection criteria lead to a trigger
efficiency of $95 \pm 2 \%$, rising to $98 \pm 2 \%$ with increasing
\scalht.

The $\gamma + \text{jets}$ sample is selected using a dedicated photon
trigger requiring a localised, large energy deposit in the ECAL with
$\Et > 150\GeV$ that satisfies loose photon identification and
isolation criteria~\cite{PAS-EGM-10-006}. The offline selection
requires $\scalht > 375\GeV$, $\alphat > 0.55$, and a single photon to
be reconstructed with $\Et > 165\GeV$, $|\eta| < 1.45$, satisfying
tight isolation criteria, and with a minimum distance to any jet of
$\Delta R > 1.0$. For these selection criteria, the photon trigger
condition is found to be fully efficient.

\subsection{Method\label{sec:method}}

The method used to estimate the non-multijet backgrounds in the signal
region relies on the use of transfer factors, which are constructed
per data control sample in bins of \scalht, \njet, and \nb. These
transfer factors are determined from simulated event samples, which
are produced as follows. The production of W and Z bosons in
association with jets is simulated with the \MADGRAPH
V5~\cite{madgraph} event generator. The production of \ttbar and
single-top quark events is generated with \POWHEG~\cite{powheg}, and
diboson events are produced with \PYTHIA6.4~\cite{pythia}. For all
simulated samples, \PYTHIA6.4 is used to describe parton showering and
hadronisation.
The description of the detector response is implemented using the
\GEANTfour~\cite{geant} package.
The simulated samples are normalised using the most accurate cross
section calculations currently available, usually with
next-to-leading-order (NLO) accuracy. To model the effects of pileup,
the simulated events are generated with a nominal distribution of pp
interactions per bunch crossing and then reweighted to match the
pileup distribution as measured in data.

Each transfer factor is defined as the ratio of expected yields as
given by simulation in a given bin of the signal region, $N_\mathrm{MC}^\text{signal}$,
and the corresponding bin of one of the control samples, $N_\mathrm{MC}^\text{control}$.
Each transfer factor is then used to extrapolate from the event yield
measured in a data control sample, $\nobs^\text{control}$,
to an expectation for the event yield in the corresponding bin of the
signal region, $\npre^\text{signal}$,
via the expression:
\begin{equation}
  \label{equ:pred-method}
  \npre^\text{signal} = \frac{N_\mathrm{MC}^\text{signal}}{N_\mathrm{MC}^\text{control}} \times \nobs^\text{control}.
\end{equation}

Two independent estimates of the irreducible background of $\znunu +
\text{jets}$ events are determined from the data control samples
comprising $\cPZ\rightarrow\mu\mu + \text{jets}$ and $\gamma +
\text{jets}$ events, both of which have similar kinematic properties
when the muons or photon are ignored~\cite{Bern:2011pa} but different
acceptances. Of the $\gamma + \text{jets}$ and $\cPZ\rightarrow\mu\mu
+ \text{jets}$ processes, the former has a larger production cross
section while the latter has kinematic properties that are more
similar to $\znunu + \text{jets}$.

The $\mu + \text{jets}$ data sample provides an estimate for the total
contribution from all other SM processes, which is dominated by \ttbar
and W-boson production. Residual contributions from single-top-quark
and diboson production are also estimated. For the category of events
satisfying $\nb \geq 2$, in which the contribution from $\znunu + \text{jets}$
events is suppressed to a negligible level, the \mj sample is also
used to estimate this small contribution rather than using the
statistically limited \mmj and \gj samples. Hence, only the \mj sample
is used to estimate the total SM background for events satisfying $\nb
\geq 2$, whereas all three data control samples are used for events
satisfying $\nb \leq 1$.

In order to maximise sensitivity to potential new physics signatures
in final states with multiple b-quark jets, a method that improves the
statistical power of the predictions from simulation, particularly for
$n_\cPqb \ge 2$, is employed. The distribution of $n_\cPqb$ is
estimated from generator-level information contained in the
simulation. The number of reconstruction-level jets matched to
underlying bottom quarks ($n_\cPqb^\text{gen}$), charm quarks
($n_\cPqc^\text{gen}$), and light-flavoured partons ($n_\cPq^\text{gen}$) per event, $N(n_\cPqb^\text{gen},n_\cPqc^\text{gen},n_\cPq^\text{gen})$, is recorded in bins of \scalht for each \njet
category.
The b-tagging efficiency, $\epsilon$, and mistag probabilities,
$f_\cPqc$ and $f_\cPq$, are also determined from simulation for
each \scalht bin and \njet category, with each quantity averaged over
jet \pt and $\eta$. Corrections are applied on a jet-by-jet
basis to both $\epsilon$, $f_\cPqc$, and $f_\cPq$ in order to
match the corresponding measurements from
data~\cite{CMS-PAS-BTV-12-001}. This information is sufficient to
predict $n_\cPqb$ and thus also determine the event yield $N(n_\cPqb)$ from simulation for a given \scalht bin and \njet category with
the expression:
\begin{equation}
  \label{equ:btag-formula}
  N(n_{\cPqb}) = \sum_{n_{\text{jet}}} \, \sum_{n_{\cPqb}}
  \left( \, N(n_{\cPqb}^{\text{gen}},
    n_{\cPqc}^{\text{gen}}, n_{\cPq}^{\text{gen}})
    \times P_\cPqb \times P_\cPqc \times P_\cPq \, \right)\, ,
\end{equation}
where $n_{\cPqb}^{\text{tag}}$,
$n_{\cPqc}^{\text{tag}}$, and $n_{\cPq}^{\text{tag}}$
are the number of times that a reconstructed b-quark jet is identified
as originating from an underlying bottom quark, charm quark, or
light-flavoured parton, respectively, and $P_\cPqb \equiv
P(n_{\cPqb}^{\text{tag}} ; n_{\cPqb}^{\text{gen}},
\epsilon)$, $P_\cPqc \equiv P(n_{\cPqc}^{\text{tag}} ;
n_{\cPqc}^{\text{gen}}, f_\cPqc)$, and $P_\cPq \equiv
P(n_{\cPq}^{\text{tag}} ; n_{\cPq}^{\text{gen}},
f_\cPq)$ are the binomial probabilities for this to happen.
The outer summation considers all
possible combinations of $n_{\cPqb}^{\text{gen}}$,
$n_{\cPqc}^{\text{gen}}$, and $n_{\cPq}^{\text{gen}}$
that satisfy $n_{\textrm{jet}} = n_{\cPqb}^{\text{gen}} +
n_{\cPqc}^{\text{gen}} + n_{\cPq}^{\text{gen}}$, while
the inner summation considers all possible combinations of
$n_{\cPqb}^{\text{tag}}$, $n_{\cPqc}^{\text{tag}}$, and
$n_{\cPq}^{\text{tag}}$ that satisfy $n_{\cPqb} =
n_{\cPqb}^{\text{tag}} + n_{\cPqc}^{\text{tag}} +
n_{\cPq}^{\text{tag}}$.

The predicted yields are found to be in good statistical agreement
with the yields obtained directly from the simulation in the bins with
a significant population. The method exploits the ability to make
precise measurements of $N(n_\cPqb^\text{gen},n_\cPqc^\text{gen},n_\cPq^\text{gen})$, $\epsilon$, $f_\cPqc$, and $f_\cPq$
independently of $n_\cPqb$, which means that event yields for a
given b-quark jet multiplicity can be predicted with a higher
statistical precision than obtained directly from simulation. Precise
measurements of $f_\cPqc$ and $f_\cPq$ are particularly important
for events with $n_\cPqb \geq 3$, which often occur in the SM
because of the presence of mistagged jets in the event. In this case,
the largest background is \ttbar, with two correctly tagged b-quark
jets and an additional mistagged jet originating from a charm quark or
light-flavoured parton.

\subsection{Systematic uncertainties on transfer factors\label{sec:syst}}

As described in Section~\ref{sec:method}, the method to estimate the
background contributions from SM processes with significant \met is
based on an extrapolation from a measurement in a control sample to a
yield expectation in the signal region. This approach aims to minimise
the sensitivity to simulation mismodelling, as many systematic biases
are expected largely to cancel in the ratios used to define the
transfer factors. However, a systematic uncertainty is assigned to
each transfer factor to account for theoretical
uncertainties~\cite{Bern:2011pa} and residual biases in the simulation
modelling of kinematics (\eg acceptances) and instrumental effects
(\eg reconstruction inefficiencies).

The magnitudes of the systematic uncertainties assigned to the
transfer factors are determined from a representative set of
closure tests in data. These tests use yields from an event category
in one of the three independent data control samples, along with the
corresponding transfer factors obtained from simulation, to predict
the yields in another event category or data control sample following
the prescription defined in Eq.~(\ref{equ:pred-method}). As stated
previously, the contamination from multijet events or any potential
signal is expected to be negligible. Therefore, the closure tests
carried out between control samples probe the properties of the
relevant SM non-multijet backgrounds.

\begin{figure*}[thbp]
  \begin{center}
    \includegraphics[width=0.48\textwidth]{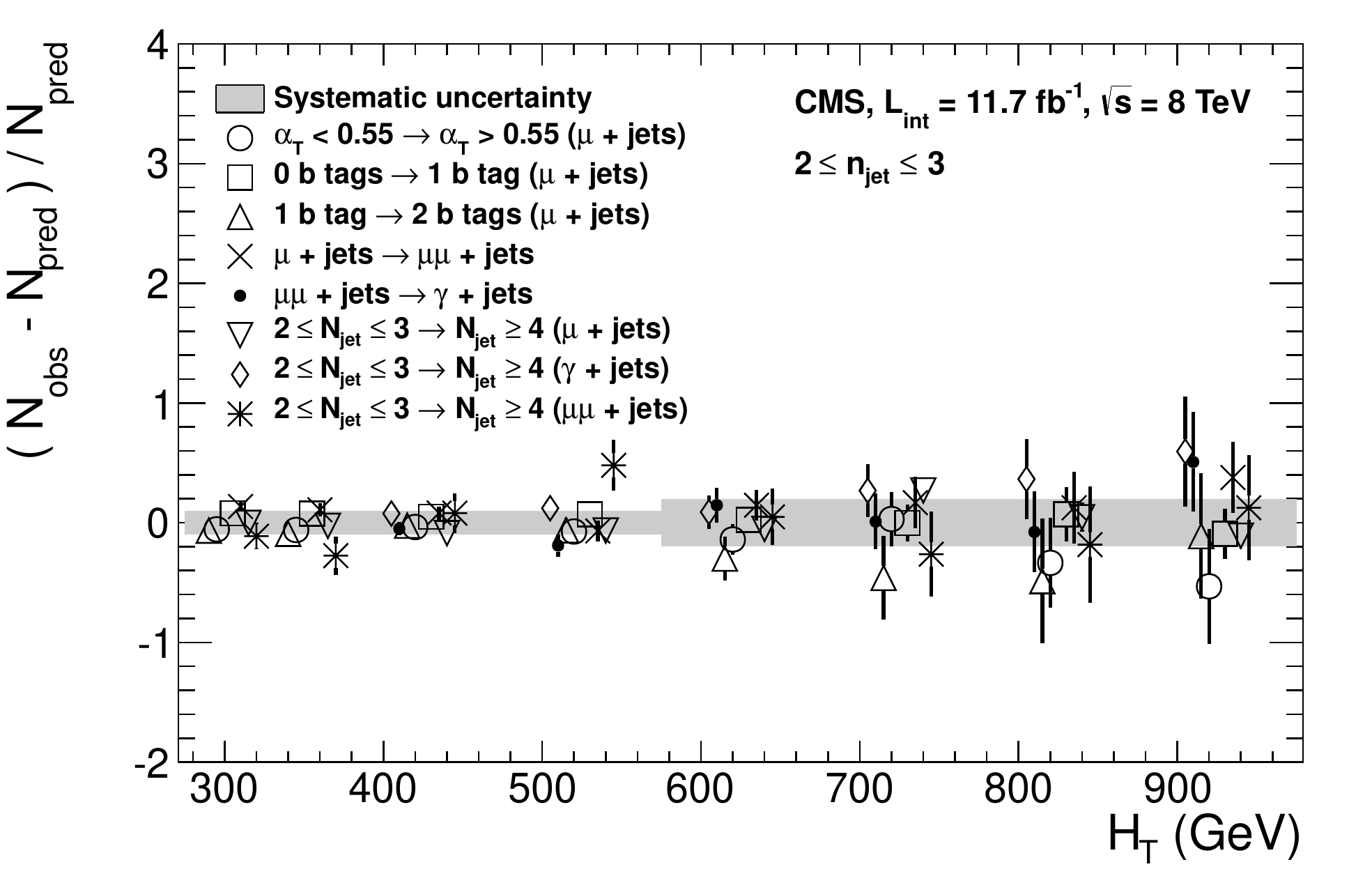}
    \includegraphics[width=0.48\textwidth]{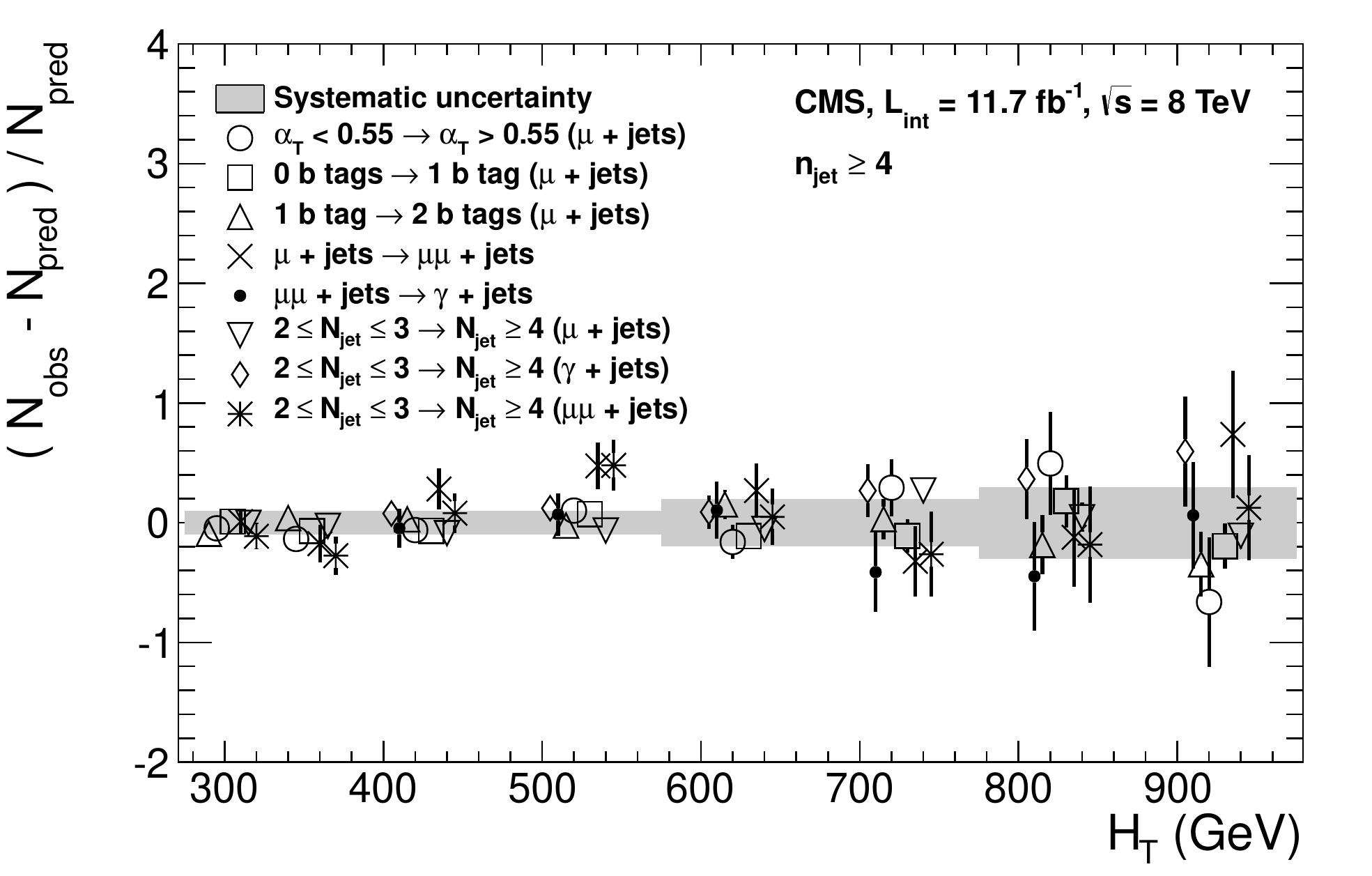} \\
    \caption{ Sets of closure tests that probe for possible
      \scalht-dependent biases associated with the transfer factors
      obtained from simulation, for the two event categories
      satisfying \njetlow (left) and \njethigh (right). Also shown are
      shaded bands that represent \scalht-dependent systematic
      uncertainties. (Colour figure online.)
    }
    \label{fig:closure-summary}
  \end{center}
\end{figure*}

Thirteen sets of closure tests are chosen to probe key ingredients of
the simulation modelling that may introduce biases in the transfer
factors. Each set comprises up to eight independent tests in bins of
\scalht. Five sets of closure tests are performed independently for
each of the two \njet categories, and a further three sets are common
to both categories, as shown in Fig.~\ref{fig:closure-summary}. For
each \njet category, the first three sets of closure tests are carried
out within the $\mu + \text{jets}$ sample, and probe the modelling of the
\alphat distribution in genuine \met events (circles), the relative
composition between $\PW+\text{jets}$ and top events (squares), and the
modelling of the reconstruction of b-quark jets (triangles),
respectively. The fourth set (crosses) addresses the modelling of the
vector boson samples by connecting the $\mu + \text{jets}$ and $\mu\mu +
\text{jets}$ control samples, with the former sample rich in $\PW+\text{jets}$ events
(and also with a significant contribution from top events) and the
latter in $\cPZ+\text{jets}$ events. The fifth set (solid bullets) deals with
the consistency between the $\cPZ\rightarrow\mu\mu + \text{jets}$ and $\gamma + \text{jets}$ samples, which are both used to provide an estimate of the
$\znunu + \text{jets}$ background.  Three further sets of closure tests
(inverted triangles, diamonds, asterisks), one per data control
sample, probe the simulation modelling of the \njet distribution.

All sets of closure tests demonstrate, given the statistical precision
of each test, that there are no significant biases or dependencies on
\scalht exhibited by the transfer factors obtained from simulation.
Table~\ref{tab:syst-fits} summarises the results obtained from
constant and linear polynomial fits to each set of closure tests for
the two \njet categories. The table also lists the best fit values and
uncertainties for the constant polynomial fits, which indicate the
level of closure averaged across the full \scalht range considered in
the analysis. All tests are either statistically compatible with zero
bias or at the level of a few percent or less. Finally,
Table~\ref{tab:syst-fits} also summarises the best fit values of the
slopes of the linear polynomial fits, which are typically of the order
$10^{-4}$, corresponding to a percent-level change per
100\GeV. However, in all cases, the best fit values are fully
compatible with zero, indicating that the level of closure is
\scalht-independent. The $\chi^{2}$ and number of degrees of freedom
(dof) of each fit are also quoted and indicate a reasonable
goodness-of-fit in all cases except one, which concerns the simulation
modelling of the \njet distribution in the \mj sample. The large
$\chi^2$ value is mainly attributable to a single outlier in the bin
$675 < \scalht < 775\GeV$ rather than any significant trend in
\scalht.

\newcommand{\testa}{$\alphat < 0.55 \,\ra\, \alphat > 0.55 \, (\mj)$}
\newcommand{\testb}{$\text{0 b tags} \,\ra\, \text{1 b tag} (\mj)$}
\newcommand{\testc}{$\text{1 b tag} \,\ra\, \text{2 b tags} (\mj)$}
\newcommand{\testd}{$\mj \,\ra\, \mmj$}
\newcommand{\teste}{$\mmj \,\ra\, \gj$}
\newcommand{\testf}{$\njetlow \,\ra\, \njethigh (\mj)$}
\newcommand{\testg}{$\njetlow \,\ra\, \njethigh (\gj)$}
\newcommand{\testh}{$\njetlow \,\ra\, \njethigh (\mmj)$}

\begin{table*}[tbhp]
  \topcaption{Results from constant and linear polynomial fits to sets of
    closure tests performed for each \njet category. The symbol
    identifies the set of closure tests in
    Fig.~\ref{fig:closure-summary}. The final four rows probe the
    simulation modelling of the \njet distribution. The $^\dag$
    indicates the fit repeated with a single outlier removed. }
  \label{tab:syst-fits}
  \centering
  \footnotesize
  \begin{tabular}{ lclcccc }
    \hline
          &        &                      & \multicolumn{2}{c}{Constant polynomial fit} & \multicolumn{2}{c}{Linear polynomial fit}                  \\
    \njet & Symbol & Set of closure tests & Constant                                    & $\chi^2$/dof & Slope                  & $\chi^2$/dof \\
          &        &                      &                                             &                 & [$10^{-4}\GeV^{-1}$] &                 \\
    \hline

     2--3\T          & $\bigcirc$         & \testa & $-0.06 \pm 0.02$           & 2.43/7 & $-1.3 \pm 2.2$           & 2.10/6 \\
     2--3            & $\square$          & \testb & $\phantom{-}0.07 \pm 0.02$ & 1.49/7 & $-1.6 \pm 1.6$           & 0.54/6 \\
     2--3            & $\bigtriangleup$   & \testc & $-0.07 \pm 0.03$           & 4.19/7 & $-2.7 \pm 3.0$           & 3.41/6 \\
     2--3            & $\times$           & \testd & $\phantom{-}0.10 \pm 0.03$ & 5.64/7 & $-1.1 \pm 2.3$           & 5.40/6 \\
     2--3            & $\bullet$          & \teste & $-0.06 \pm 0.04$           & 5.93/5 & $\phantom{-}4.2 \pm 4.3$ & 4.98/4 \\
     $\geq$4\T       & $\bigcirc$         & \testa & $-0.05 \pm 0.03$           & 9.58/7 & $\phantom{-}3.0 \pm 2.9$ & 8.47/6 \\
     $\geq$4         & $\square$          & \testb & $-0.03 \pm 0.03$           & 5.88/7 & $-1.0 \pm 1.9$           & 5.59/6 \\
     $\geq$4         & $\bigtriangleup$   & \testc & $-0.02 \pm 0.03$           & 7.35/7 & $\phantom{-}1.1 \pm 2.2$ & 7.08/6 \\
     $\geq$4         & $\times$           & \testd & $\phantom{-}0.08 \pm 0.07$ & 12.9/7 & $\phantom{-}4.8 \pm 4.3$ & 11.7/6 \\
     $\geq$4         & $\bullet$          & \teste & $-0.03 \pm 0.10$           & 2.85/5 & $-4.0 \pm 7.0$           & 2.52/4 \\
     $\geq$2\T          & $\bigtriangledown$ & \testf & $-0.03 \pm 0.02$           & 17.3/7 & $\phantom{-}0.0 \pm 1.0$ & 17.3/6 \\
     $\geq$2$\,\,^\dag$ & $\bigtriangledown$ & \testf & $-0.04 \pm 0.01$           & 6.10/6 & $-1.4 \pm 1.1$           & 4.46/5 \\
     $\geq$2            & $\lozenge$         & \testg & $\phantom{-}0.12 \pm 0.05$ & 2.42/5 & $\phantom{-}6.0 \pm 4.7$ & 0.77/4 \\
     $\geq$2            & $*$                & \testh & $-0.04 \pm 0.07$           & 9.76/7 & $\phantom{-}4.9 \pm 4.4$ & 8.51/6 \\

     \hline
  \end{tabular}
\end{table*}

Once it is established that no significant bias or trend is observed
for any set of closure tests, uncorrelated systematic uncertainties on
the transfer factors are determined for five independent regions in
\scalht: 275--325, 325--375, 375--575, 575--775, and $\geq775\GeV$.
Conservative estimates for the systematic uncertainties are based on
the variance in the level of closure for all individual tests,
weighted according to the statistical uncertainties associated with
each test, within a given \scalht region. This procedure yields
estimates of 10\% (10\%), 10\% (10\%), 10\% (10\%), 20\% (20\%), and
20\% (30\%) for the five \scalht regions defined above for events
satisfying \njetlow (\njethigh), as indicated by the shaded bands in
Fig.~\ref{fig:closure-summary}.

The effect on the transfer factors of uncertainties related to the
modelling of b-quark jets in simulation is studied in detail. After
correcting the b-tagging efficiency and mistag probability determined
in simulation for residual differences as measured in data, the
corresponding uncertainties on these corrections are propagated to the
transfer factors. In addition, several robustness tests are performed,
\eg treating c-quark jets as b-quark jets.
While the absolute yields ($N_\mathrm{MC}^\text{signal}$ and
$N_\mathrm{MC}^\text{control}$) are susceptible to systematic biases,
the transfer factors are not, because changes to
$N_\mathrm{MC}^\text{signal}$ and $N_\mathrm{MC}^\text{control}$ are
strongly correlated. The relative change in the transfer factors is
found to be negligible, at the sub-percent level. Hence, the
aforementioned \scalht-dependent systematic uncertainties are also
used for each \nb category and are treated as uncorrelated among \nb
categories.

\section{Estimating the multijet background\label{sec:multijet}}

The contribution from multijet events is expected to be negligible, at
or below the percent-level relative to the yields expected from
non-multijet backgrounds, even for the most inclusive definition of
the signal region, defined by $\scalht > 275\GeV$, $\alphat > 0.55$,
and no requirement on \njet or \nb. The expected yield is further
suppressed to $\ll 1$ event with the application of more stringent
thresholds on any of the variables \scalht, \njet, or \nb.

Any potential contamination from multijet events via the effects
described in Sections~\ref{sec:alphat} and \ref{sec:signal} can be
estimated by exploiting the \scalht dependence of the ratio of events
with a value of \alphat above and below some threshold,
$\rat(\scalht)$. This dependence on \scalht is modelled as a falling
exponential function, $\rat(\scalht) = A\re^{-k
  H_\mathrm{T}}$~\cite{RA1Paper}, where the parameters $A$ and $k$ are
the normalisation and decay constant parameters, respectively. The
exponential model is validated in a multijet-enriched data sideband,
defined by the event selection criteria for the signal region,
described in Section~\ref{sec:signal}, but with the requirement
$\mht/\met > 1.25$. A measurement of the decay constant $k$ is made in
a further multijet-enriched sample defined by the event selection
criteria for the signal region but with the requirement $\alphat <
0.55$.

The estimate of the multijet contamination in the signal region for a
given \scalht bin is determined from the product of the ratio \RaT, as
given by the exponential model, and the yield in a data control sample
defined by the event selection for the signal region but with the
requirement $\alphat < 0.55$. This event sample is recorded with a set
of trigger conditions that require only \scalht to be above the same
thresholds as used by the signal region triggers listed in
Table~\ref{tab:triggers}.

Further details on the exponential model and its use in the likelihood
model are found in Section~\ref{sec:results}.
\section{Confronting data with the SM-only hypothesis\label{sec:results}}

For a given category of events satisfying requirements on both \njet
and \nb, a likelihood model of the observations in multiple data
samples is used to obtain a consistent prediction of the SM
backgrounds and to test for the presence of a variety of signal
models. It is written as:

\begin{align}
  \label{likelihood}
  L_{\njet,\,\nb} \; & = \; L_\text{SR} \times L_{\mu} \times
  L_{\mu\mu} \times L_{\gamma} \, ; & (0 \leq \nb \leq 1) \\
  L_{\njet,\,\nb} \; & = \; L_\text{SR} \times L_{\mu} \, , & (\nb
  \geq 2)
\end{align}

where $L_\text{SR}$ describes the yields in the eight \scalht bins of
the signal region where exactly \njet jets and \nb b-quark jets are
required. In each bin of \scalht, the observation is modelled as a
Poisson-distributed variable about the sum of the SM expectation and a
potential signal contribution (assumed to be zero in the following
discussion), where the SM expectation is the sum of the multijet and
non-multijet components. The non-multijet component is related to the
expected yields in the \mj, \mmj, and \gj control samples via the
transfer factors derived from simulation, as described in
Section~\ref{sec:method}. The likelihood functions $L_\mu$,
$L_{\mu\mu}$, and $L_\gamma$ describe the yields in the \scalht bins
of the \mj, \mmj, and \gj control samples in the same category of
\njet and \nb as the signal region. For the category of events
satisfying $\nb \geq 2$, only the $\mu + \text{jets}$ control sample
is used in the likelihood to determine the total contribution from all
non-multijet SM backgrounds in the signal region.
The estimate of the contribution from multijet events in a given
\scalht bin of the signal region relies on the exponential model
$\rat(\scalht) = A\re^{-k H_\mathrm{T}}$, as described in
Section~\ref{sec:multijet}.
The systematic uncertainties (10--30\%) associated with the transfer
factors, discussed in Section~\ref{sec:syst}, are accommodated in the
likelihood function by a nuisance parameter per transfer factor. The
measurements of these parameters are assumed to follow a log-normal
distribution.

In order to test the compatibility of the observed yields with the
expectations from only SM processes, the likelihood function is
maximised over all fit parameters. For each of the eight categories of
events defined by requirements on \njet and \nb, the goodness-of-fit
of the SM-only hypothesis is determined by considering simultaneously
up to eight bins in \scalht from the signal region and up to 22 bins
from the three control samples. No significant tension is observed in
the signal and control regions, which are well described by the SM
hypothesis. The p-values obtained are found to be uniformly
distributed, with a minimum observed value of
0.1. Table~\ref{tab:ensemble} summarises the observed yields and fit
results in bins of \scalht for events in the signal region categorised
according to \njet and \nb.

\begin{figure*}[tbhp]
  \begin{center}
    \includegraphics[width=0.45\textwidth]{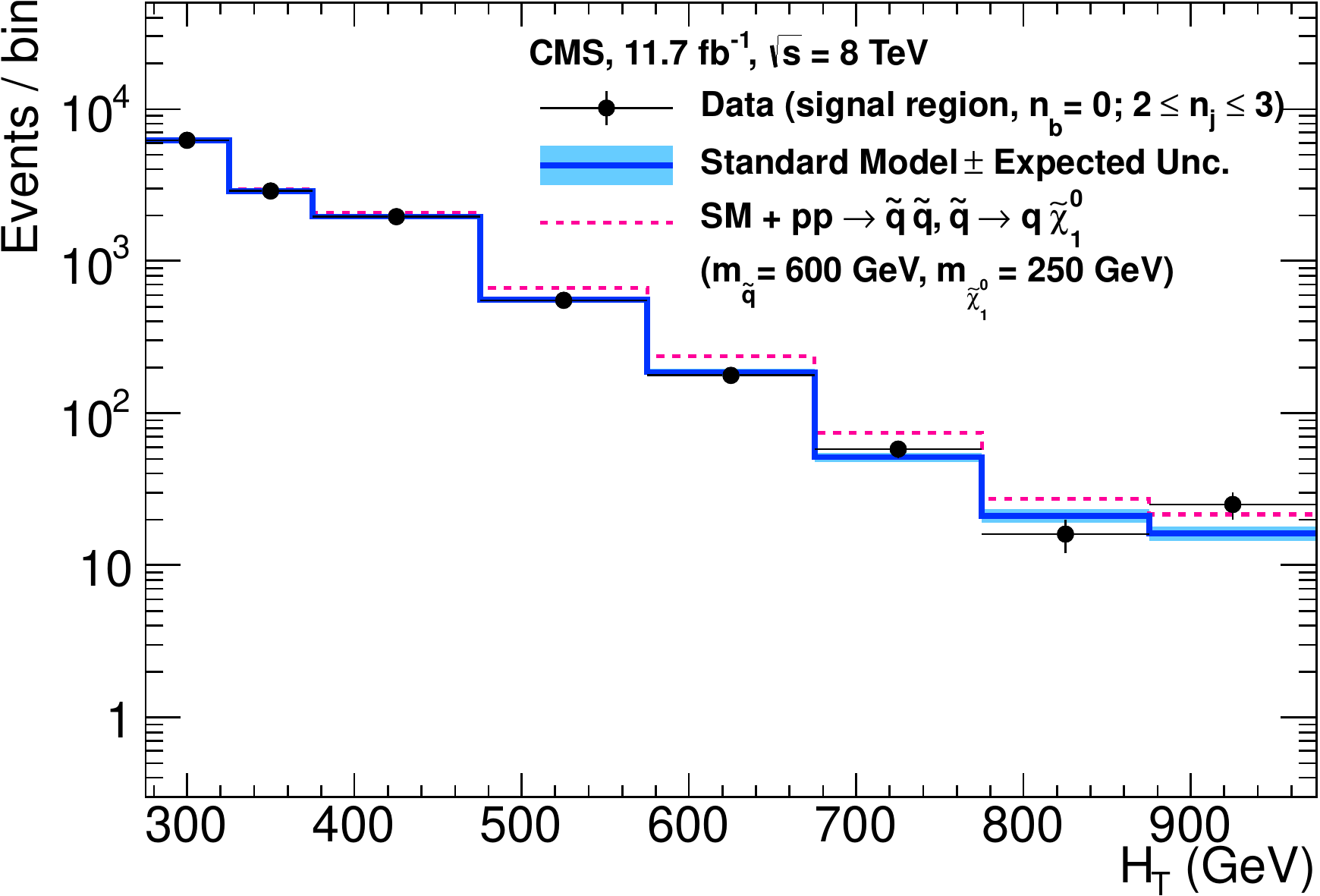}
    \includegraphics[width=0.45\textwidth]{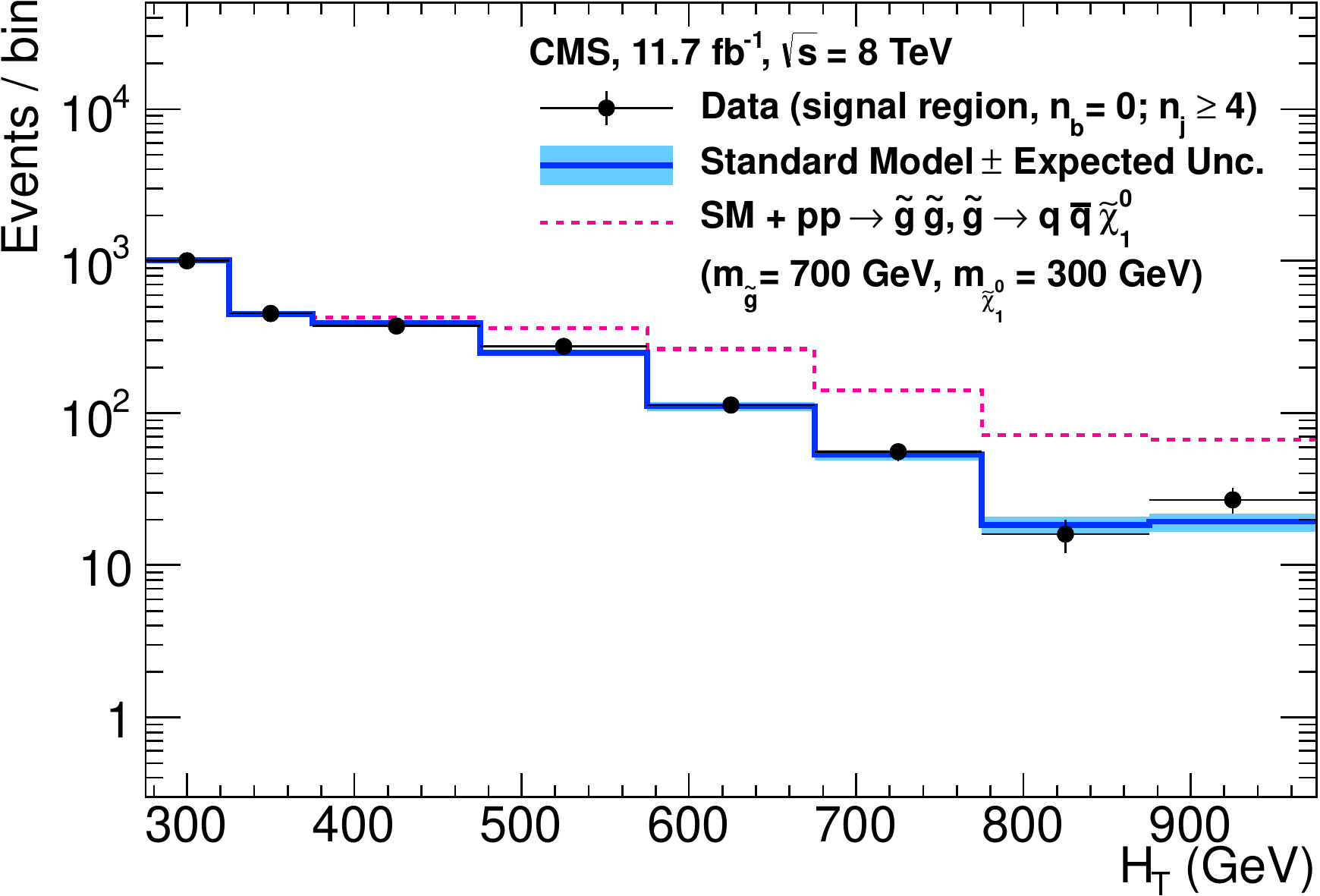} \\
    \caption{\label{fig:best-fit-0b} Event yields observed in data
      (solid circles) and SM expectations with their associated
      uncertainties (solid lines with bands) in bins of \scalht for
      the signal region when requiring exactly zero b-quark jets and
      \njetlow (left) or \njethigh (right). For illustration only, the
      expectations for the reference mass points of the signal models
      \Ttwo (left, red dashed line) and \Tone (right, red dashed line)
      are superimposed on the SM-only expectations. (Colour figure online.)}
  \end{center}
\end{figure*}

\begin{figure*}[tbhp]
  \begin{center}
    \includegraphics[width=0.45\textwidth]{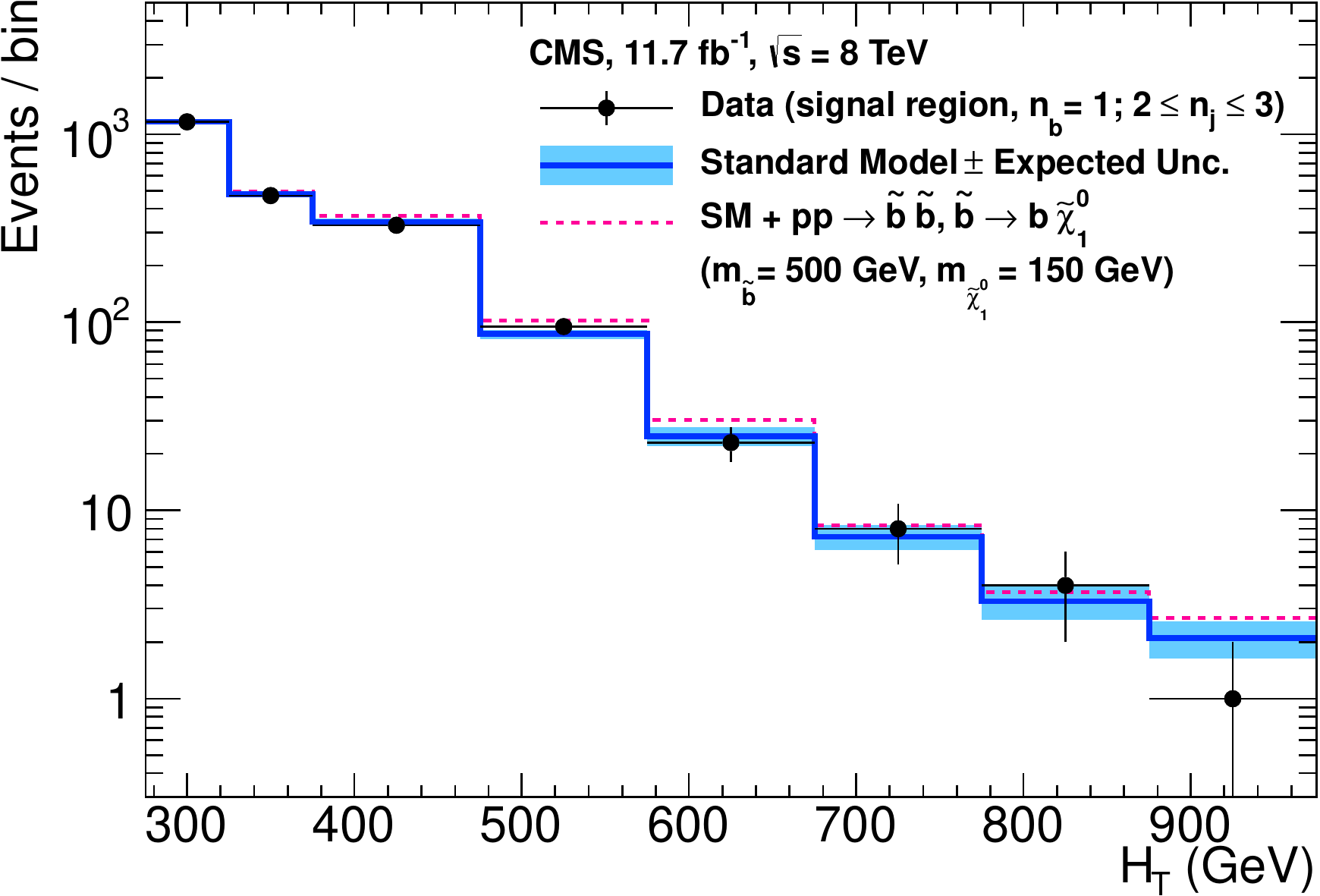}
    \includegraphics[width=0.45\textwidth]{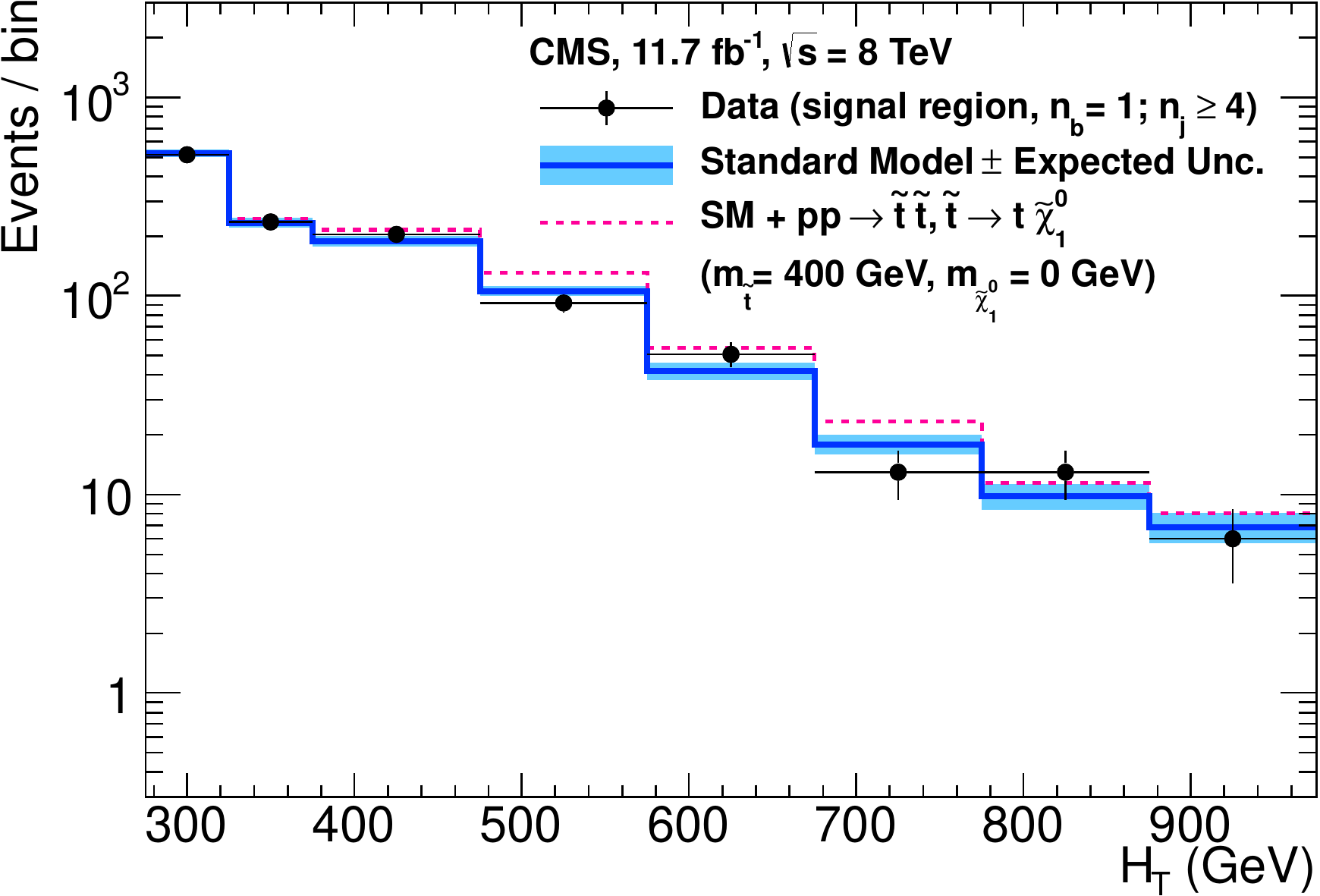} \\
    \caption{\label{fig:best-fit-1b} As for
      Fig.~\ref{fig:best-fit-0b}, but requiring exactly one b-quark
      jet and \njetlow (left) or \njethigh (right). Example signal
      yields are for the reference mass points of the signal models
      \TtwoBB (left, red dashed line) and \TtwoTT (right, red dashed
      line). (Colour figure online.) }
  \end{center}
\end{figure*}

\begin{figure*}[tbhp]
  \begin{center}
    \includegraphics[width=0.45\textwidth]{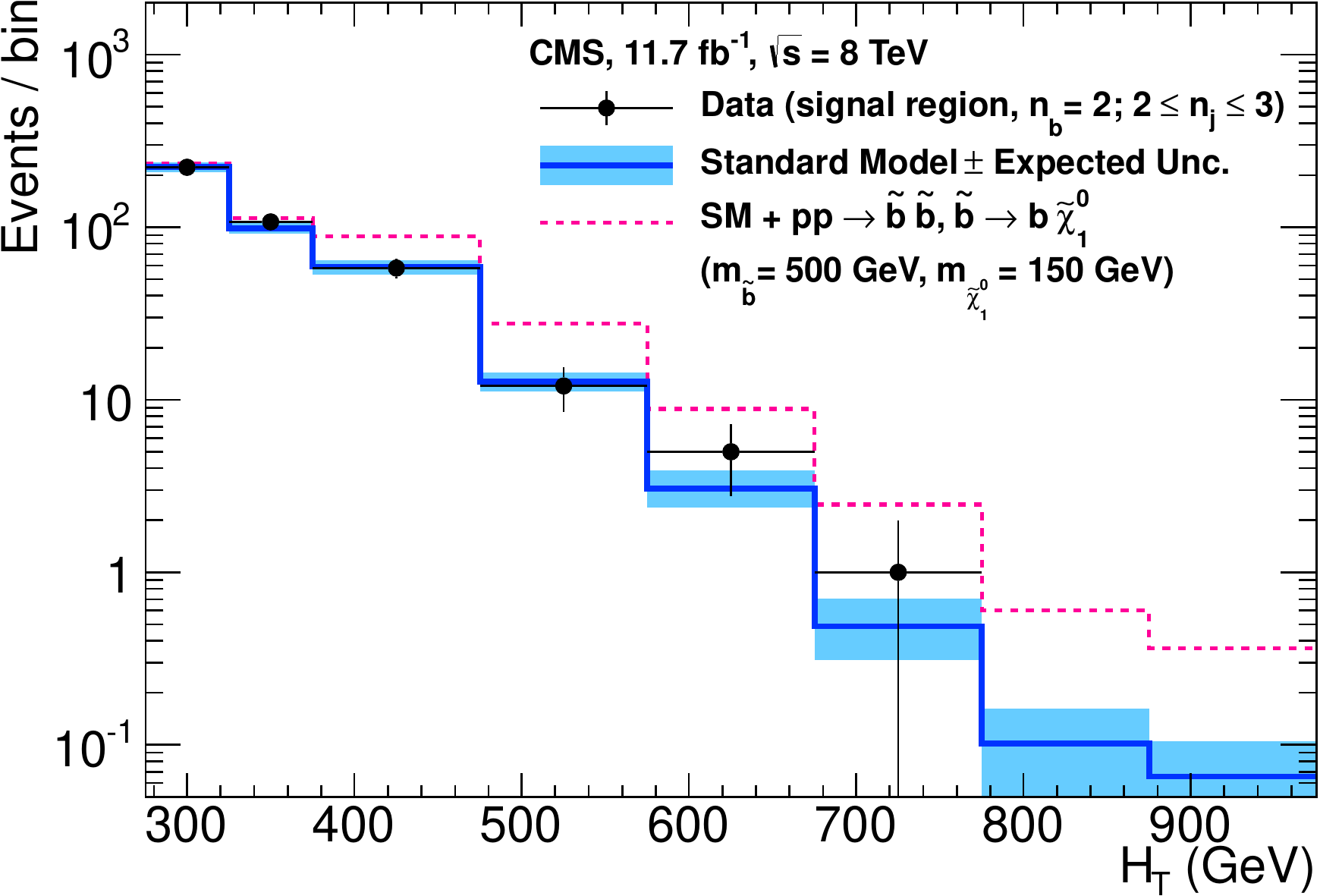}
    \includegraphics[width=0.45\textwidth]{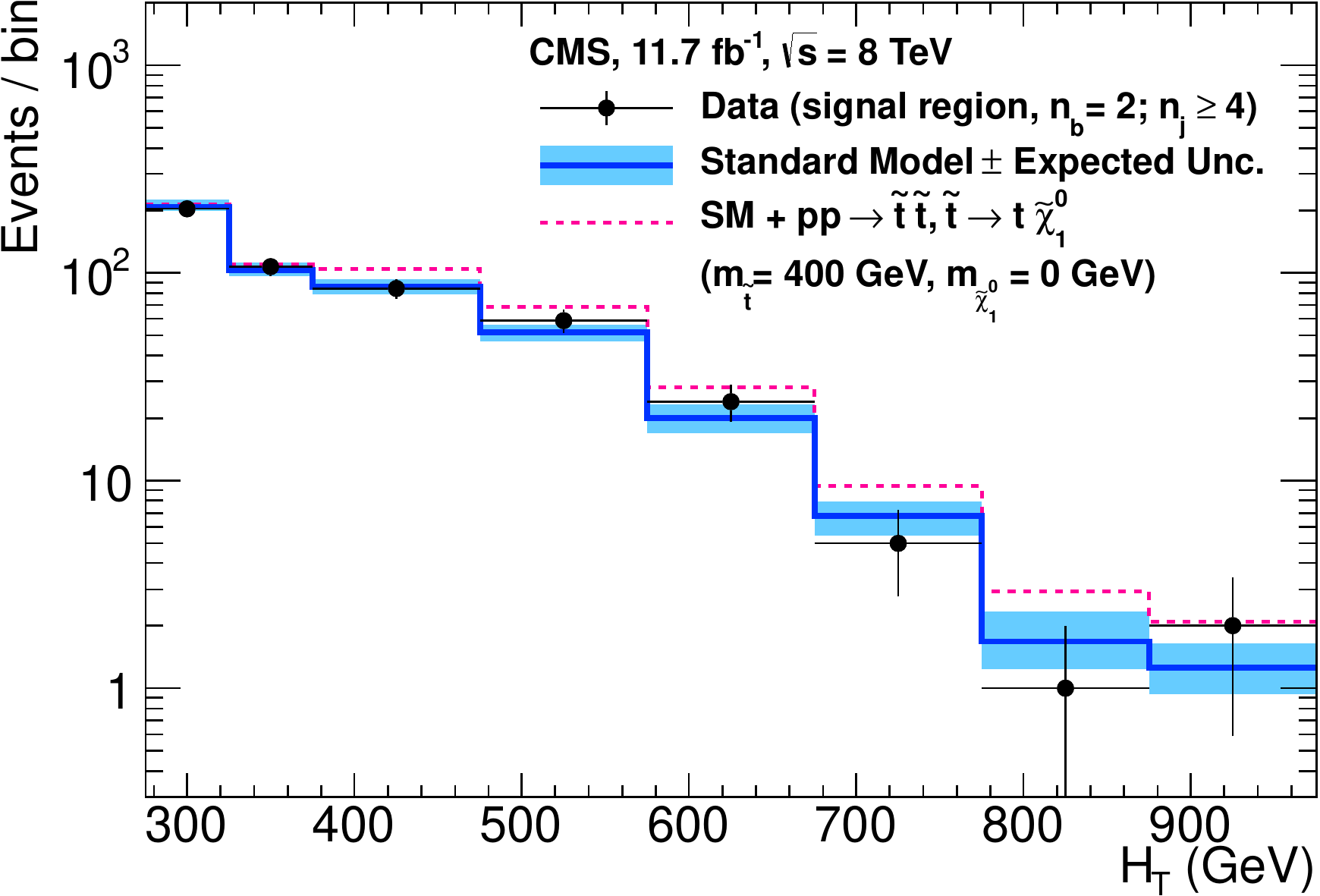} \\
    \caption{\label{fig:best-fit-2b} As for
      Fig.~\ref{fig:best-fit-0b}, but requiring exactly two b-quark
      jets and \njetlow (left) or \njethigh (right). Example signal
      yields are for the reference mass points of the signal models
      \TtwoBB (left, red dashed line) and \TtwoTT (right, red dashed
      line). (Colour figure online.) }
  \end{center}
\end{figure*}

\begin{figure*}[tbhp]
  \begin{center}
    \includegraphics[width=0.45\textwidth]{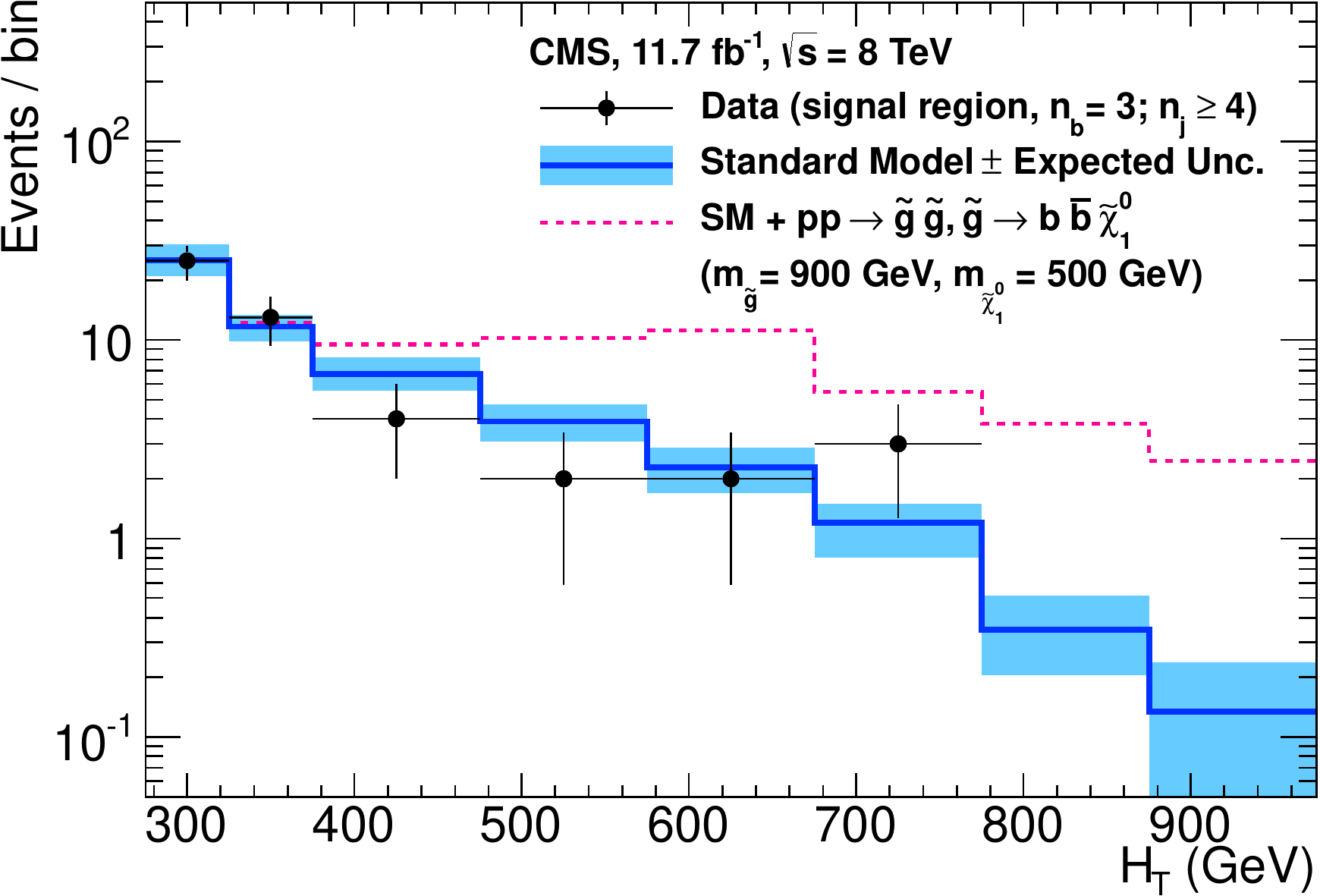}
    \includegraphics[width=0.45\textwidth]{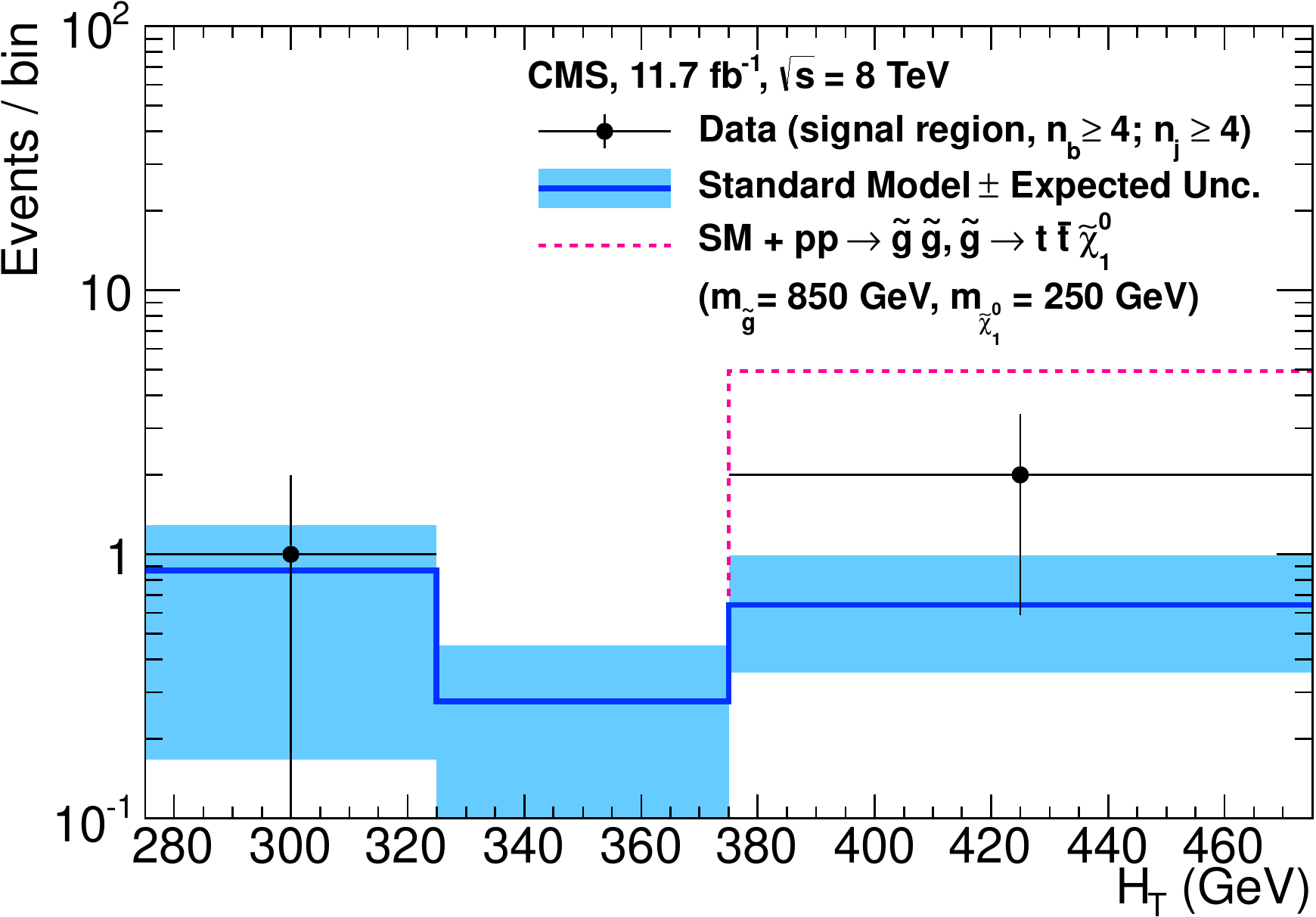} \\
    \caption{\label{fig:best-fit-ge3b} As for
      Fig.~\ref{fig:best-fit-0b}, but requiring \njethigh and exactly
      three (left) or at least four (right) b-quark jets. Example
      signal yields are for the reference mass points of the signal
      models \ToneBBBB (left, red dashed line) and \ToneTTTT (right,
      red dashed line). (Colour figure online.) }
  \end{center}
\end{figure*}

\begin{figure*}[tbhp]
  \begin{center}
    \includegraphics[width=0.45\textwidth]{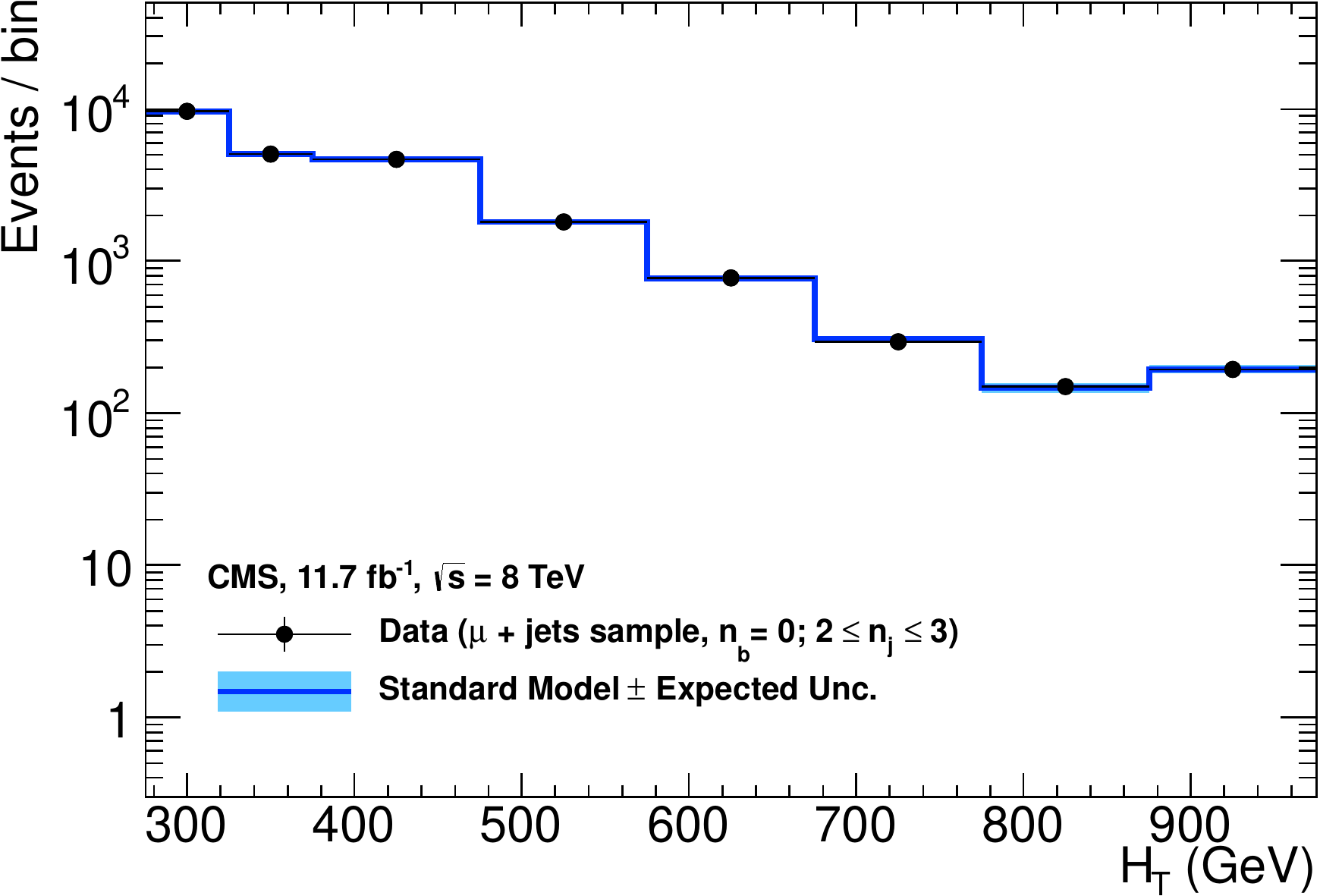}
    \includegraphics[width=0.45\textwidth]{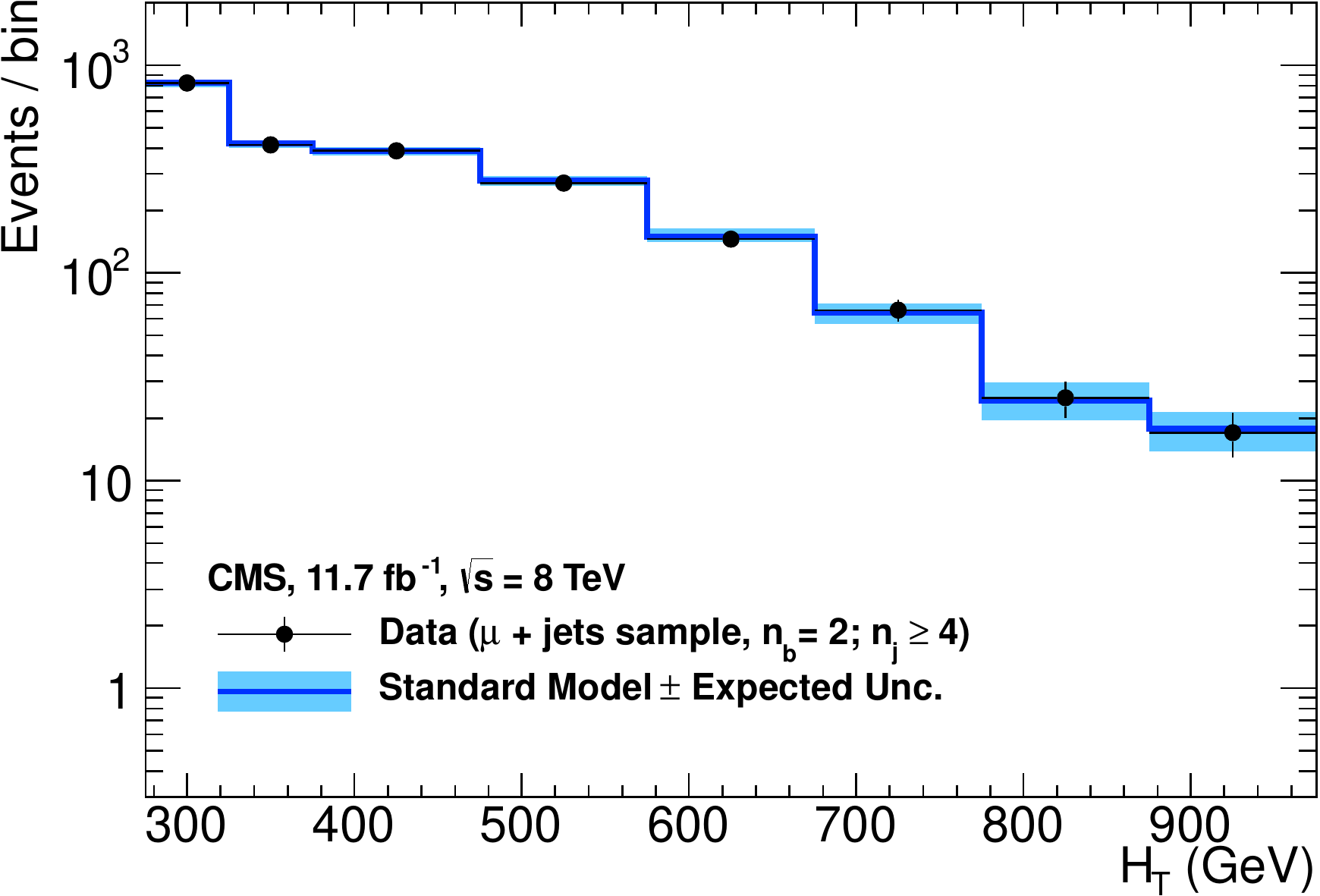} \\
    \includegraphics[width=0.45\textwidth]{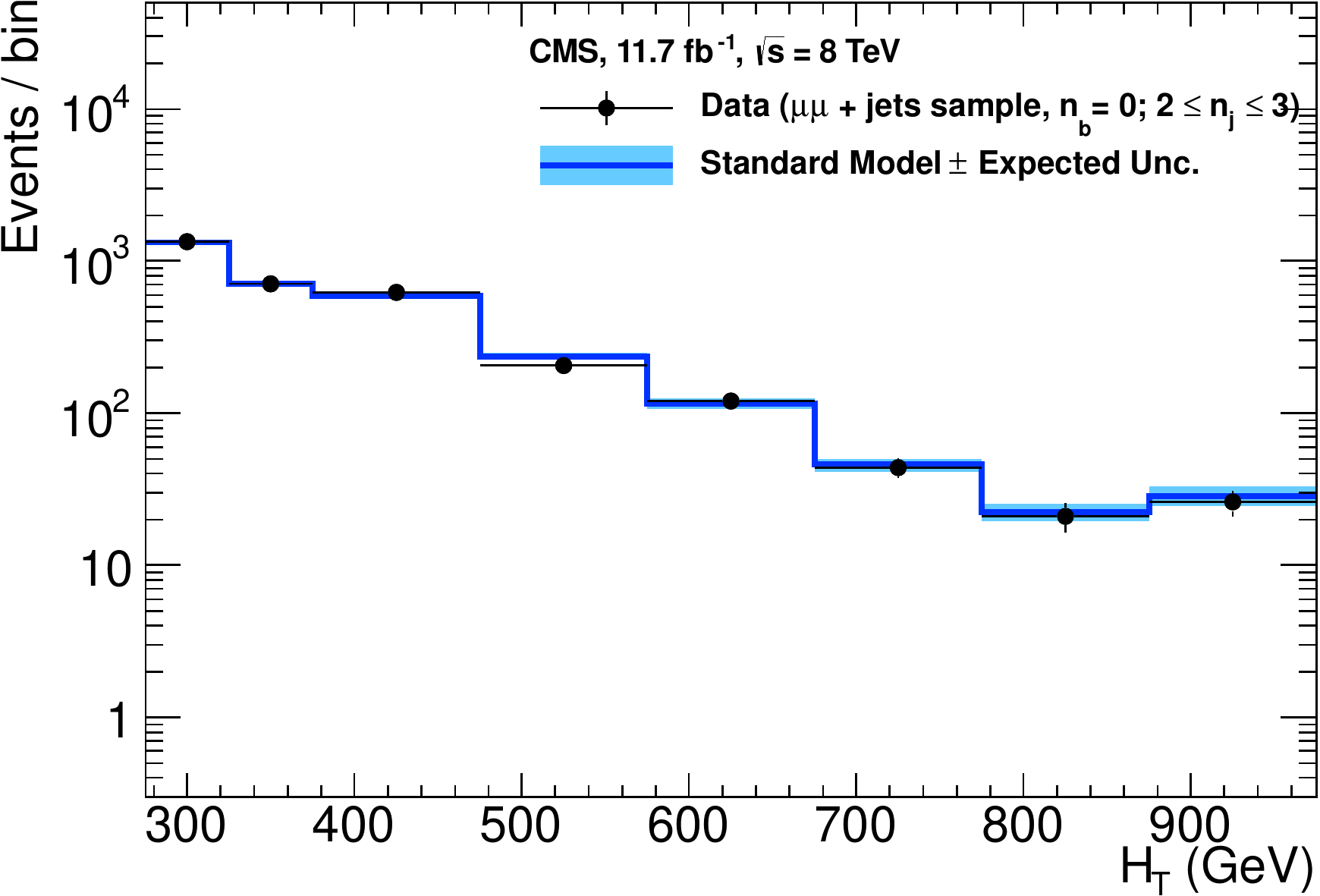}
    \includegraphics[width=0.45\textwidth]{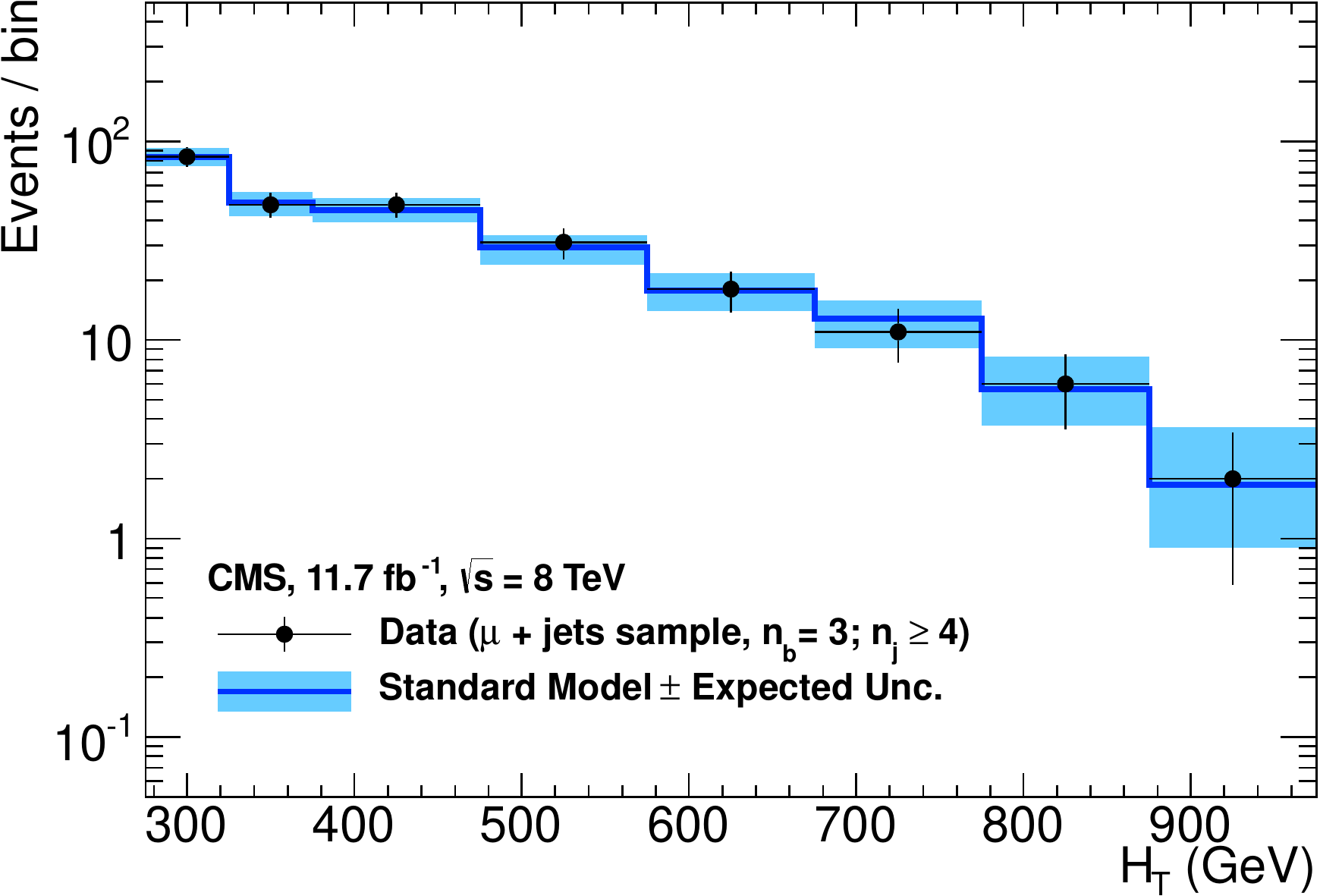} \\
    \includegraphics[width=0.45\textwidth]{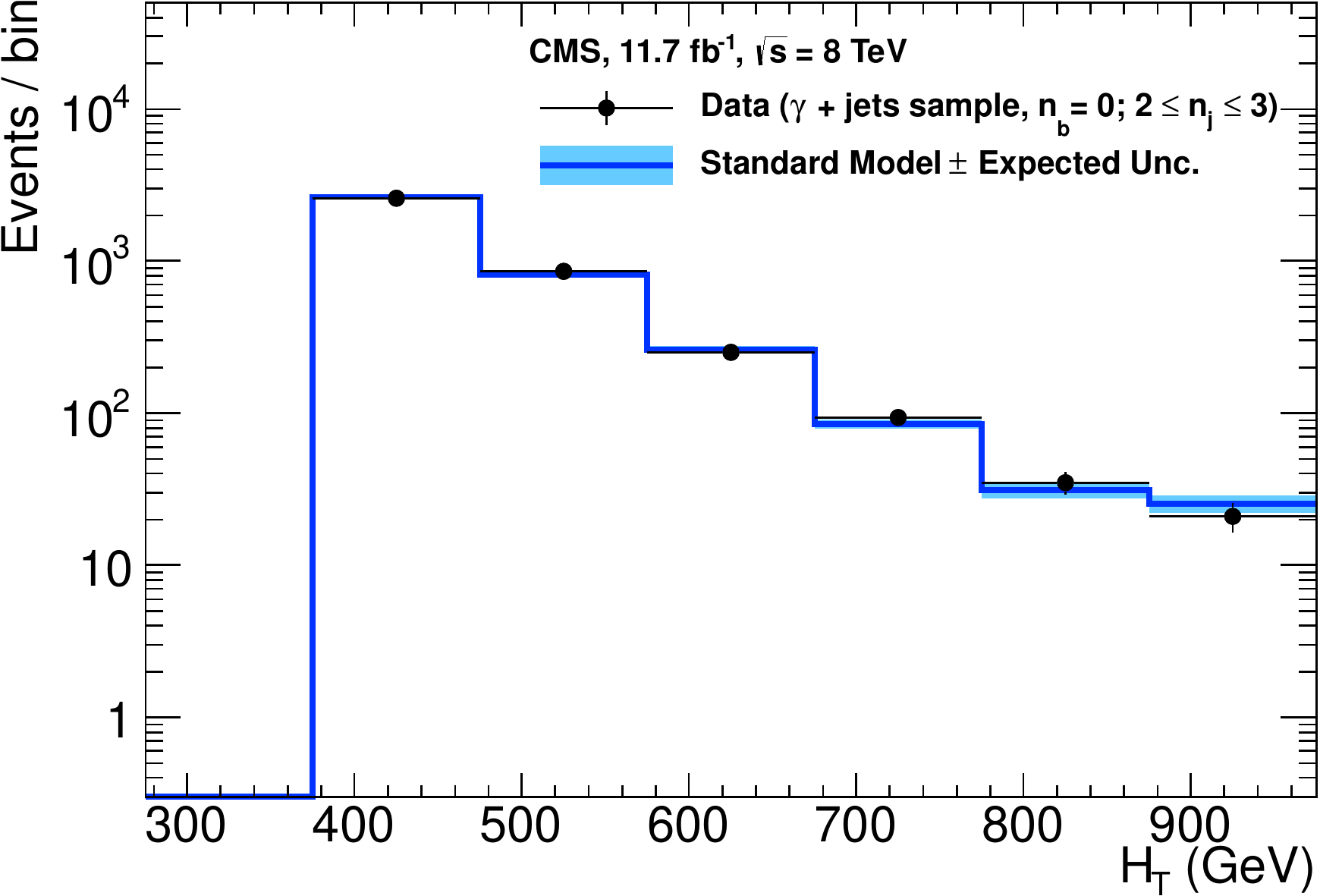}
    \includegraphics[width=0.45\textwidth]{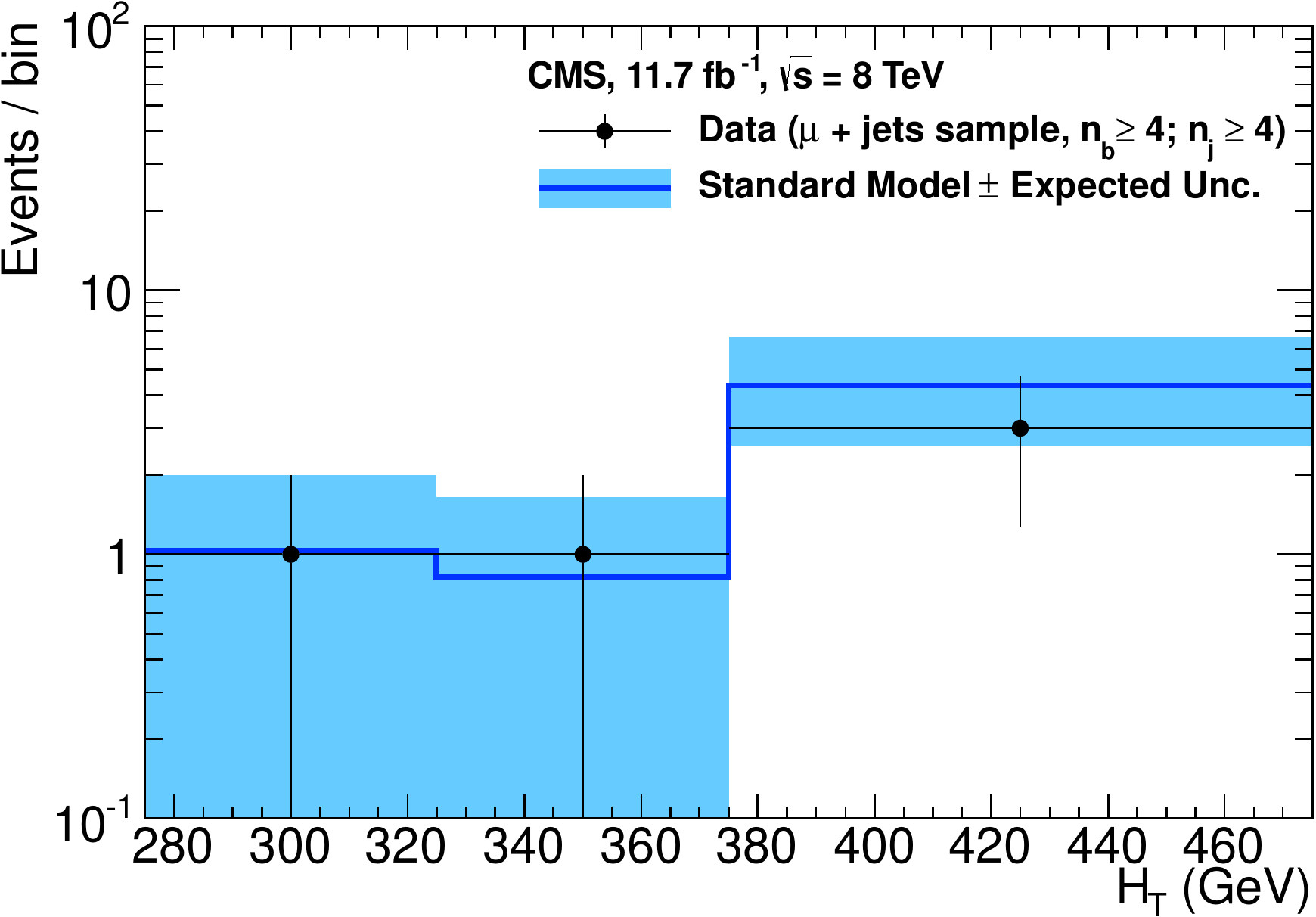} \\
    \caption{\label{fig:best-fit-control} Event yields observed in
      data (solid circles) and SM expectations with their associated
      uncertainties (solid lines with bands) in bins of \scalht for:
      the \mj (top left), \mmj (middle left), and \gj (bottom left)
      control samples when requiring \njetlow and exactly zero b-quark
      jets; and the \mj control sample when requiring \njethigh and
      exactly two (top right), three (middle right), or at least four
      (bottom right) b-quark jets. (Colour figure online.)}
  \end{center}
\end{figure*}

Comparisons of the observed yields and the SM expectations in bins of
\scalht for events categorised according to \njet and containing
exactly zero, one, or two b-quark jets are shown in
Figs.~\ref{fig:best-fit-0b}, \ref{fig:best-fit-1b}, and
\ref{fig:best-fit-2b}, respectively. Similarly,
Fig.~\ref{fig:best-fit-ge3b} shows the \scalht-binned observed yields
and SM expectations for events satisfying \njethigh and $\nb = 3$
(left) or $\nb \geq 4$ (right). For illustration,
Figs.~\ref{fig:best-fit-0b}--\ref{fig:best-fit-ge3b} include the
expected yields from various reference models, as defined in
Table~\ref{tab:simplified-models}. Figure~\ref{fig:best-fit-control}
(left column) shows the observed yields and SM expectations in the
\scalht bins of the \mj, \mmj, and \gj control samples for events
satisfying \njetlow and $\nb = 0$. Figure~\ref{fig:best-fit-control}
(right column) shows the observed yields and SM expectations in the
\scalht bins of the \mj sample for events satisfying \njethigh and
$\nb = 2$, $\nb = 3$, or $\nb \geq 4$.

The maximum-likelihood values for the decay constant and normalisation
parameters, $k$ and $A$, of the exponential model for the multijet
background are obtained independently for each of the eight event
categories. The value of the nuisance parameter $k$ is constrained via
a measurement in a multijet-enriched data sideband, as described in
Section~\ref{sec:multijet}. No constraint is applied to the
normalisation term. In the nominal fit, the maximum-likelihood value
of the normalisation parameter for each event category is found to be
compatible with zero within uncertainties. Furthermore, the expected
yields obtained from an alternate fit, in which the normalisation
parameters are fixed to zero, agree well with those obtained from the
nominal fit.

\begin{table*}[tbhp]
  \topcaption{Event yields observed in data and fit results with
    their associated uncertainties in bins of \scalht for events in
    the signal region that are categorised according to \njet and
    \nb. The final $\scalht > 375\GeV$ bin is inclusive for the
    $\njet \geq 4$ and $\nb \geq 4$ category.
  }
  \label{tab:ensemble}
  \footnotesize
  \centering
  \begin{tabular}{ lllllllllll }
    \hline
    \multicolumn{3}{c}{} & \multicolumn{8}{c}{\scalht bin [\GeVns{}]} \\ [0.5ex]
    \cline{4-11}
     & \njet   & \nb     & 275--325             & 325--375             & 375--475                    & 475--575             & 575--675             & 675--775             & 775--875             & 875--$\infty$        \\ [0.5ex]
    \hline
    SM\T   & 2--3 & 0    & 6235$^{+100}_{-67}$  & 2900$^{+60}_{-54}$   & 1955$^{+34}_{-39}$          & 558$^{+14}_{-15}$    & 186$^{+11}_{-10}$    & 51.3$^{+3.4}_{-3.8}$ & 21.2$^{+2.3}_{-2.2}$ & 16.1$^{+1.7}_{-1.7}$ \\
    Data\B & 2--3 & 0    & 6232                 & 2904                 & 1965                        & 552                  & 177                  & 58                   & 16                   & 25                   \\
    SM\T   & 2--3 & 1    & 1162$^{+37}_{-29}$   & 481$^{+18}_{-19}$    & 341$^{+15}_{-16}$           & 86.7$^{+4.2}_{-5.6}$ & 24.8$^{+2.8}_{-2.7}$ & 7.2$^{+1.1}_{-1.0}$  & 3.3$^{+0.7}_{-0.7}$  & 2.1$^{+0.5}_{-0.5}$  \\
    Data\B & 2--3 & 1    & 1164                 & 473                  & 329                         & 95                   & 23                   & 8                    & 4                    & 1                    \\
    SM\T   & 2--3 & 2    & 224$^{+15}_{-14}$    & 98.2$^{+8.4}_{-6.4}$ & 59.0$^{+5.2}_{-6.0}$        & 12.8$^{+1.6}_{-1.6}$ & 3.0$^{+0.9}_{-0.7}$  & 0.5$^{+0.2}_{-0.2}$  & 0.1$^{+0.1}_{-0.1}$  & 0.1$^{+0.1}_{-0.1}$  \\
    Data\B & 2--3 & 2    & 222                  & 107                  & 58                          & 12                   & 5                    & 1                    & 0                    & 0                    \\
    SM\T   & $\geq$4 & 0    & 1010$^{+34}_{-24}$   & 447$^{+19}_{-16}$    & 390$^{+19}_{-15}$           & 250$^{+12}_{-11}$    & 111$^{+9}_{-7}$      & 53.3$^{+4.3}_{-4.3}$ & 18.5$^{+2.4}_{-2.4}$ & 19.4$^{+2.5}_{-2.7}$ \\
    Data\B & $\geq$4 & 0    & 1009                 & 452                  & 375                         & 274                  & 113                  & 56                   & 16                   & 27                   \\
    SM\T   & $\geq$4 & 1    & 521$^{+25}_{-17}$    & 232$^{+15}_{-12}$    & 188$^{+12}_{-11}$           & 106$^{+6}_{-6}$      & 42.1$^{+4.1}_{-4.4}$ & 17.9$^{+2.2}_{-2.0}$ & 9.8$^{+1.5}_{-1.4}$  & 6.8$^{+1.2}_{-1.1}$  \\
    Data\B & $\geq$4 & 1    & 515                  & 236                  & 204                         & 92                   & 51                   & 13                   & 13                   & 6                    \\
    SM\T   & $\geq$4 & 2    & 208$^{+17}_{-9}$     & 103$^{+9}_{-7}$      & 85.9$^{+7.2}_{-6.9}$        & 51.7$^{+4.6}_{-4.7}$ & 19.9$^{+3.4}_{-3.0}$ & 6.8$^{+1.2}_{-1.3}$  & 1.7$^{+0.7}_{-0.4}$  & 1.3$^{+0.4}_{-0.3}$  \\
    Data\B & $\geq$4 & 2    & 204                  & 107                  & 84                          & 59                   & 24                   & 5                    & 1                    & 2                    \\
    SM\T   & $\geq$4 & 3    & 25.3$^{+5.0}_{-4.2}$ & 11.7$^{+1.7}_{-1.8}$ & 6.7$^{+1.4}_{-1.2}$         & 3.9$^{+0.8}_{-0.8}$  & 2.3$^{+0.6}_{-0.6}$  & 1.2$^{+0.3}_{-0.4}$  & 0.3$^{+0.2}_{-0.1}$  & 0.1$^{+0.1}_{-0.1}$  \\
    Data\B & $\geq$4 & 3    & 25                   & 13                   & 4                           & 2                    & 2                    & 3                    & 0                    & 0                    \\
    SM\T   & $\geq$4 & $\geq$4 & 0.9$^{+0.4}_{-0.7}$  & 0.3$^{+0.2}_{-0.2}$  & 0.6$^{+0.3}_{-0.3}$ & --                   & --                   & --                   & --                   & --                   \\
    Data\B & $\geq$4 & $\geq$4 & 1                    & 0                    & 2                   & --                   & --                   & --                   & --                   & --                   \\
    \hline
  \end{tabular}
\end{table*}

\section{Interpretation of the results\label{sec:interpretation}}

Limits are set in the parent sparticle and LSP mass parameter space of
the simplified models listed in Table~\ref{tab:simplified-models}. The
\cls method~\cite{read,junk} is used to compute the limits, with the
one-sided (LHC-style) profile likelihood ratio as the test
statistic~\cite{Cowan:2010js}.  The sampling distributions for the
test statistic are built by generating pseudo-data from the likelihood
function, using the respective maximum-likelihood values of the
nuisance parameters under the SM background-only and
signal-plus-background hypotheses. Signal contributions in each of the
data samples are considered, though the only significant contribution
occurs in the signal region and not the control
samples. Table~\ref{tab:categories} specifies the event categories,
defined in terms of \njet and \nb, used to provide interpretations in
the different simplified models.

\begin{table}[tbhp]
  \topcaption{A summary of the event categories used to provide an
    interpretation in the various simplified models considered in this
    paper.
  }
  \label{tab:categories}
  \centering
  \begin{tabular}{ lrr }
    \hline
    Model     & \njet   & \nb           \\ [0.5ex]
    \hline
    \Ttwo     & 2--3    & 0             \\
    \TtwoBB   & 2--3    & 1, 2          \\
    \TtwoTT   & $\geq$4 & 1, 2          \\
    \Tone     & $\geq$4 & 0             \\
    \ToneBBBB & $\geq$4 & 2, 3, $\geq$4 \\
    \ToneTTTT & $\geq$4 & 2, 3, $\geq$4 \\
    \hline
  \end{tabular}
\end{table}

Event samples for the simplified models are generated at leading order
with \PYTHIA6.4~\cite{pythia}. Inclusive, process-dependent, NLO
calculations of SUSY production cross sections, with
next-to-leading-logarithmic (NLL) corrections, are obtained with the
program \PROSPINO~\cite{Beenakker:1996ch, PhysRevD.80.095004,
  PhysRevLett.102.111802, PhysRevD.80.095004, 1126-6708-2009-12-041,
  doi:10.1142/S0217751X11053560, susy-nlo-nll}. The samples are
generated using the CTEQ6L1~\cite{Pumplin:2002vw} PDFs.  The
distribution of the number of pp interactions per bunch crossing for
the simulated samples matches that observed in data.

Various experimental uncertainties on the expected signal yield are
considered for each interpretation. Signal acceptance in the kinematic
region defined by $0 < \mparent - \mlsp < 175\gev$ or $\mparent <
300\gev$ is due in part to the presence of initial-state
radiation. Given the large associated uncertainties from simulation
for this kinematic region, no interpretation is provided. Otherwise,
the experimental systematic uncertainties are determined for each
point in the mass parameter space of each simplified model. Models are
categorised according to the mass splitting between the parent
sparticle and the LSP, with those satisfying $175 < \mparent - \mlsp <
350\GeV$ deemed to be characterised by a compressed spectrum.
For a given category of model, \ie with a compressed spectrum or
otherwise (as defined above), the systematic uncertainties are
relatively stable throughout the mass plane, thus a single
conservative value is considered for each category.

Estimates of the various systematic uncertainties for models with a
compressed spectrum or otherwise are summarised in
Tables~\ref{tab:sms-syst-near} and \ref{tab:sms-syst-far},
respectively. Contributions from the analysis selection are dominated
by uncertainties on the PDFs, jet energy scale (JES), and modelling of
the efficiency and mistag probability of b-quark jets in simulation.
The total systematic uncertainties provided in the tables also account
for the uncertainty of 4.4\% on the luminosity measurement~\cite{lumi}
and contributions from other event selection criteria, such as: the
trigger conditions; the removal of events containing isolated muons,
electrons, or photons; and filters to suppress classes of rare,
pathological events, as described in Section~\ref{sec:signal}. Each of
these individual contributions is below 4\%. The total systematic
uncertainty on the expected signal yield for the various simplified
models is found to be in the range 12\%--23\% and is accounted for
with a nuisance parameter, the measurement of which is assumed to
follow a lognormal distribution.

\begin{table}[tbhp]
  \topcaption{Estimates of the dominant systematic uncertainties (\%),
    defined in the text, on
    the  analysis efficiency for various simplified models that are
    characterised by a small mass splitting (\ie compressed spectrum)
    between the parent sparticle and LSP. The totals also
    reflect contributions from additional systematic uncertainties
    described in the text. The region $\mparent - \mlsp < 350\GeV$ is
    kinematically forbidden for the \ToneTTTT model.
  }
  \label{tab:sms-syst-near}
  \centering
  \begin{tabular}{ lrrrrrr }
    \hline
    Model     & \Ttwo & \TtwoBB & \TtwoTT & \Tone & \ToneBBBB & \ToneTTTT \\
    \hline
    PDF       & 10.0  & 10.0    & 10.0    & 10.0  & 10.0      & --        \\
    JES       & 4.1   & 4.8     & 6.5     & 5.6   & 7.3       & --        \\
    b-tagging & 2.4   & 2.2     & 0.8     & 3.1   & 2.7       & --        \\
    Total     & 12.9  & 13.1    & 13.9    & 13.9  & 14.5      & --        \\
    \hline
  \end{tabular}
\end{table}

\begin{table}[tbhp]
  \topcaption{Estimates of the dominant systematic uncertainties (\%),
    defined in the text, on
    the analysis efficiency for various simplified models that are
    characterised by a large mass splitting between the parent
    sparticle and LSP. The totals also reflect contributions
    from additional systematic uncertainties described in the
    text.
  }
  \label{tab:sms-syst-far}
  \centering
  \begin{tabular}{ lrrrrrr }
    \hline
    Model     & \Ttwo & \TtwoBB & \TtwoTT & \Tone & \ToneBBBB & \ToneTTTT \\
    \hline
    PDF       & 10.0  & 10.0    & 10.0    & 10.0  & 10.0      & 10.0      \\
    JES       & 1.1   & 0.9     & 3.5     & 0.8   & 1.5       & 0.5       \\
    b-tagging & 5.8   & 2.7     & 1.6     & 6.6   & 10.1      & 19.4      \\
    Total     & 13.4  & 12.3    & 12.9    & 14.0  & 16.0      & 23.0      \\
    \hline
  \end{tabular}
\end{table}

\begin{figure*}[tbhp]
  \begin{center}
    \includegraphics[width=0.45\textwidth]{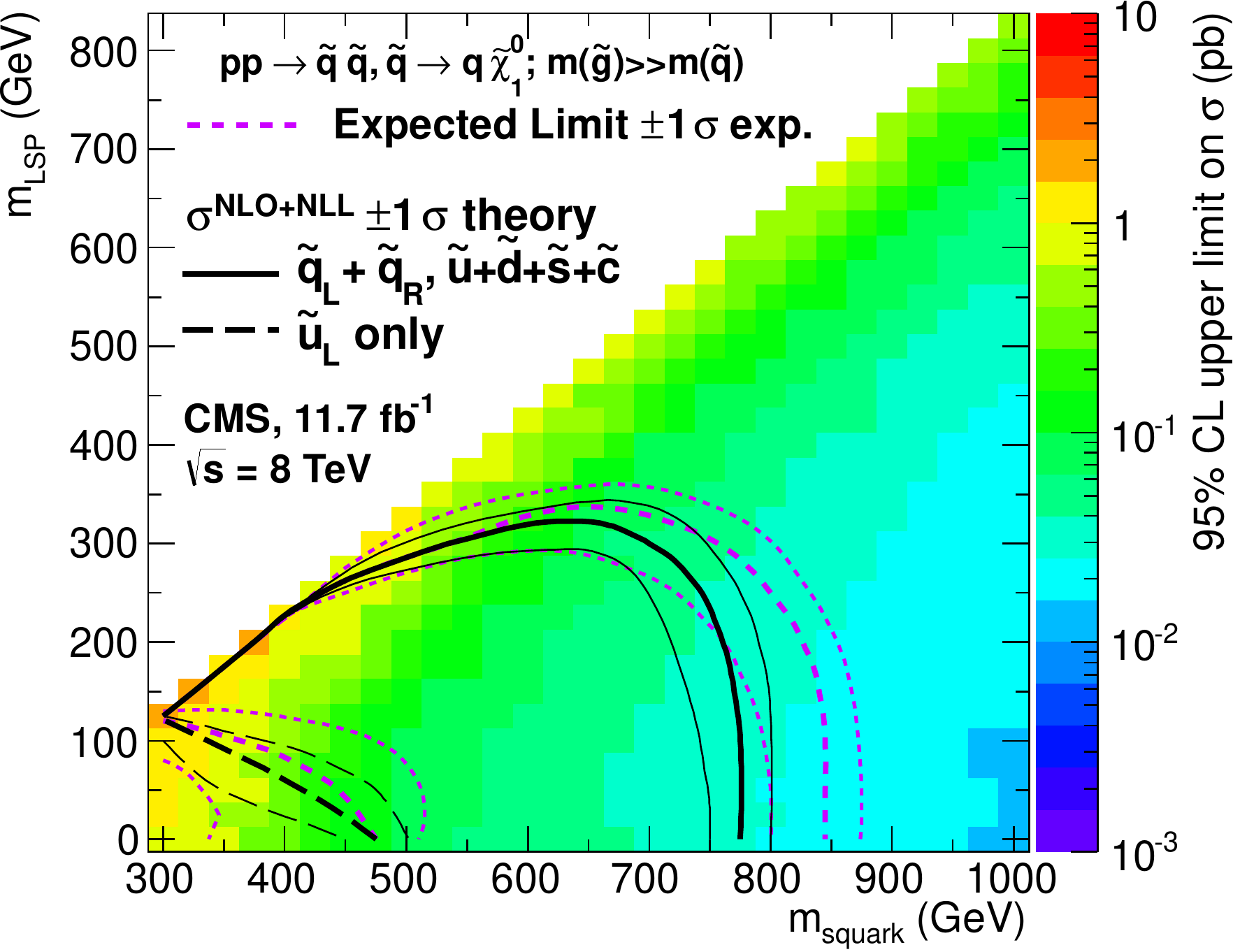} \,
    \includegraphics[width=0.45\textwidth]{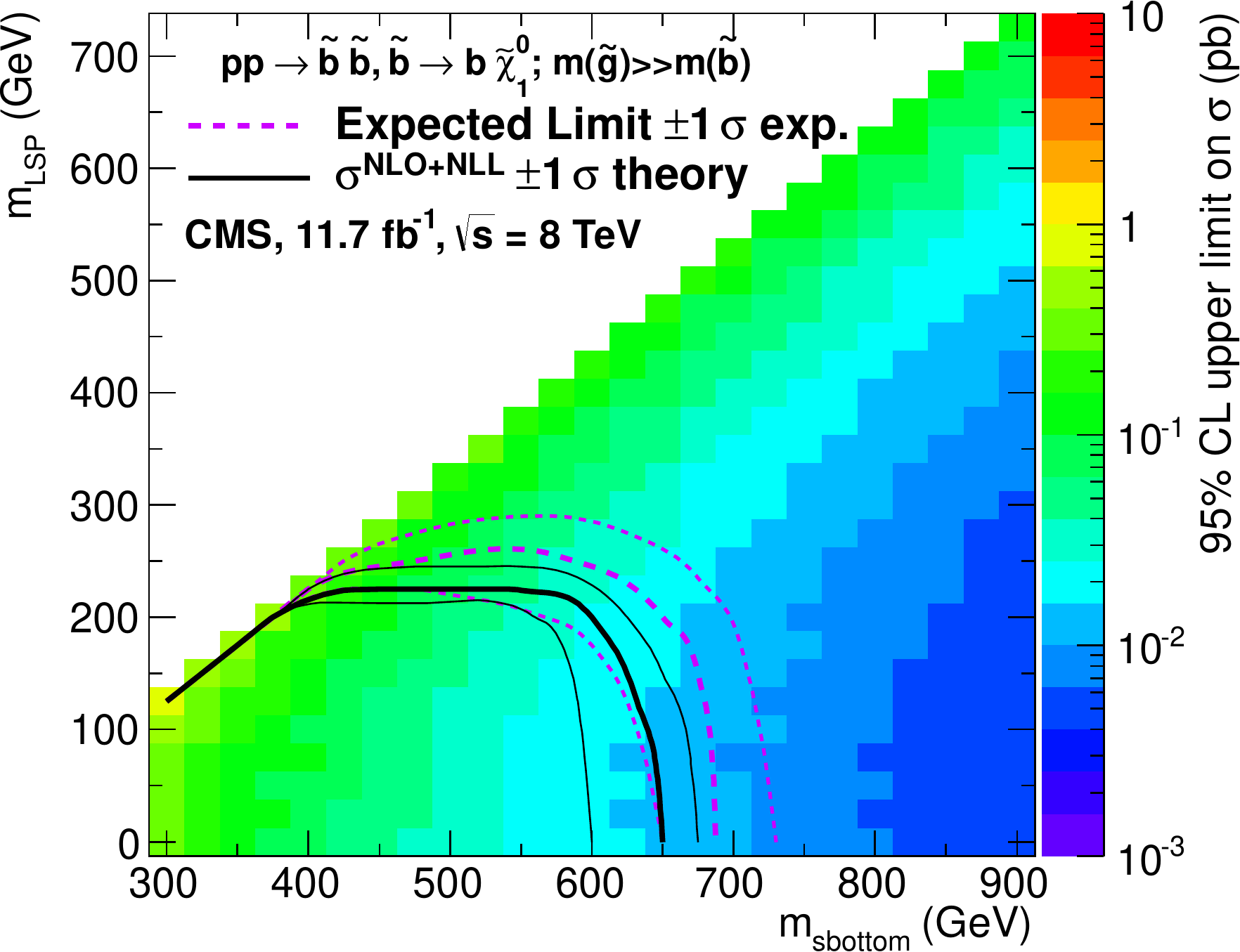} \\
    \includegraphics[width=0.45\textwidth]{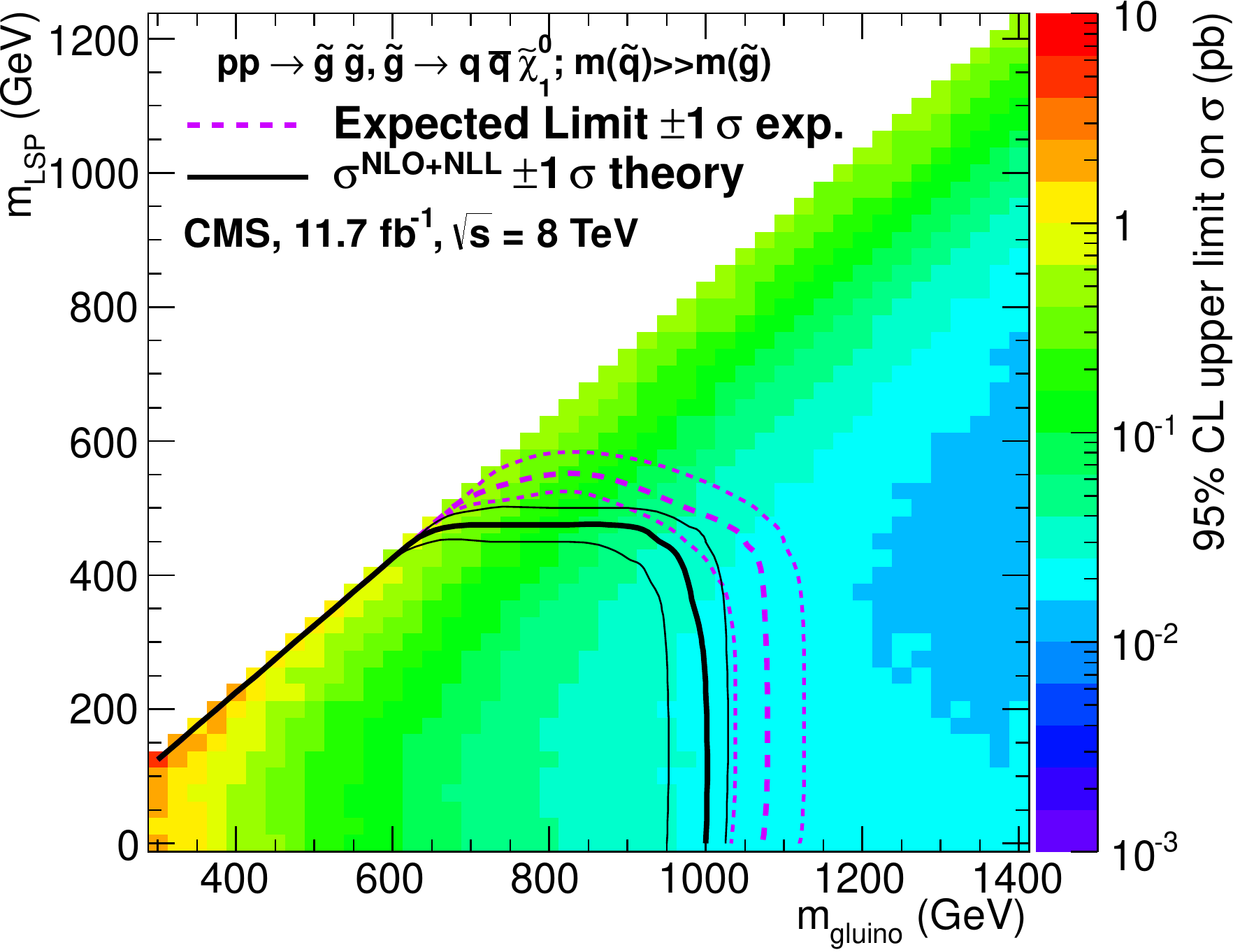} \\
    \includegraphics[width=0.45\textwidth]{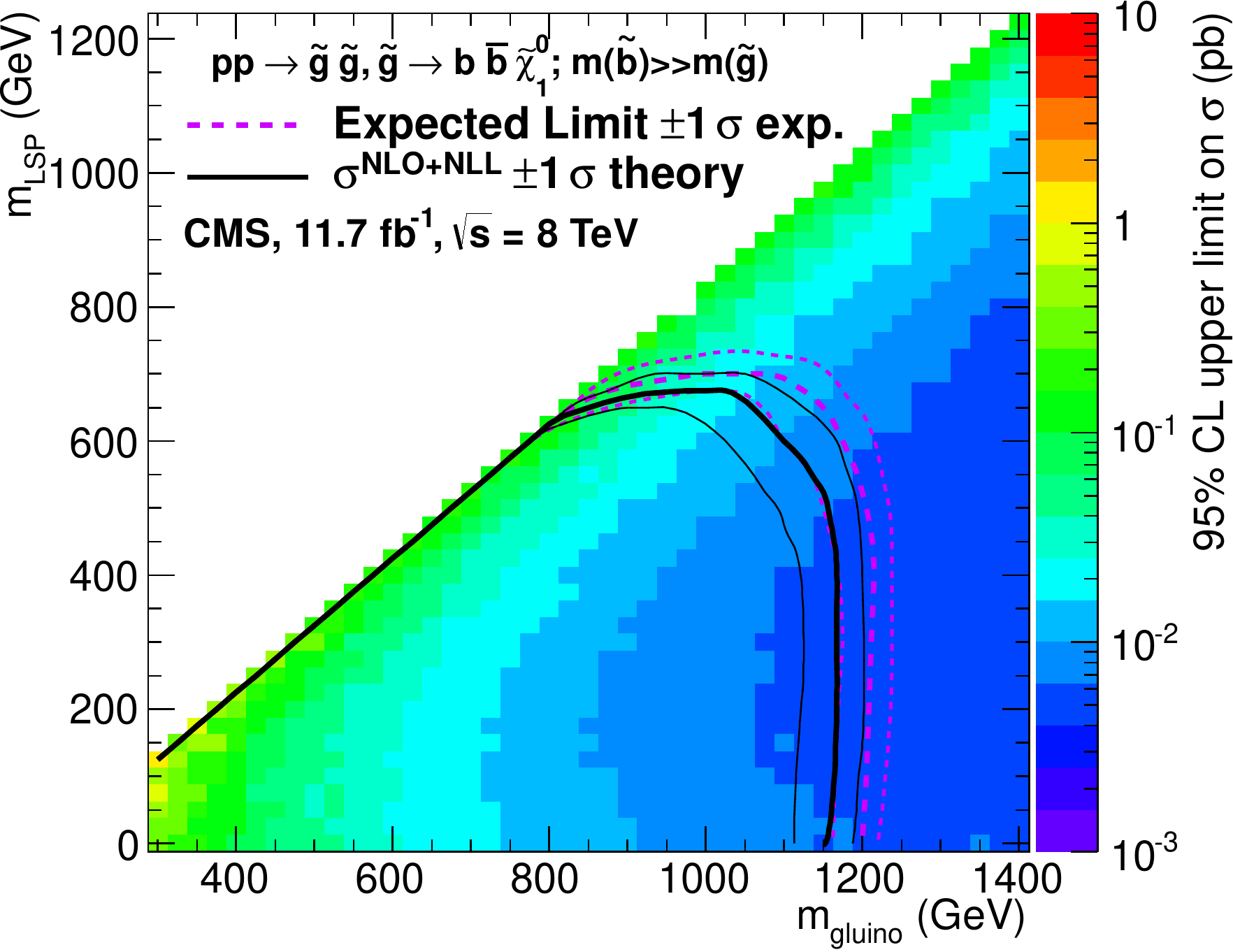} \,
    \includegraphics[width=0.45\textwidth]{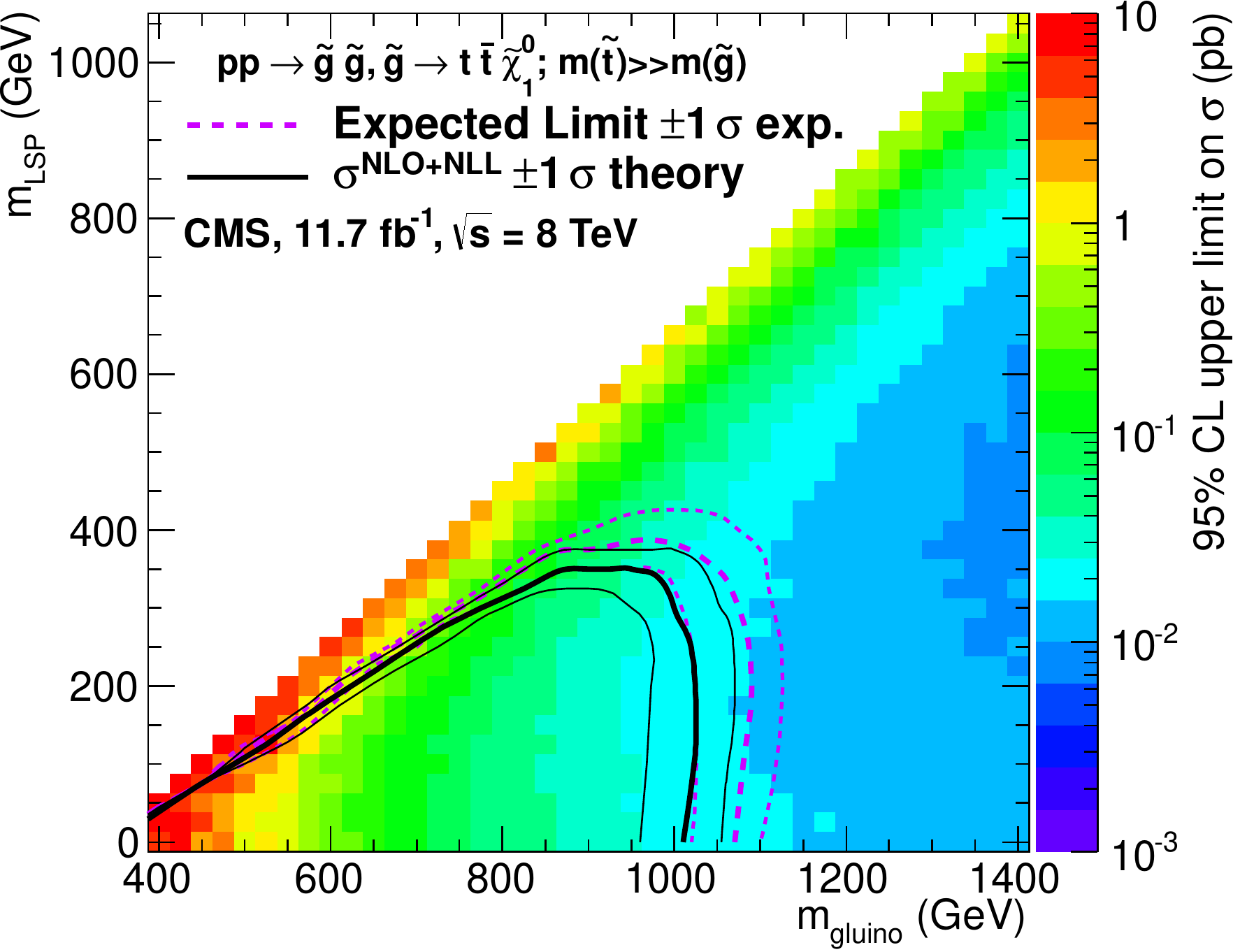} \\
    \caption{Observed upper limit on the production cross section at
      95\% CL (indicated by the colour scale) as a function of the
      parent and LSP sparticle masses for simplified models involving:
      the direct pair production of eight first- and second-generation
      squarks with degenerate masses or only a single light squark
      (\Ttwo, top left); the direct pair production of bottom squarks
      (\TtwoBB, top right); and pair-produced gluinos followed by the
      decay of each gluino to the LSP and pairs of first- and
      second-generation quarks (\Tone, middle), bottom quarks
      (\ToneBBBB, bottom left), or top quarks (\ToneTTTT, bottom
      right).  The black solid (or dashed) thick line indicates the
      observed exclusion assuming NLO+NLL SUSY production cross
      section. The black solid (or dashed) thin lines represent the
      observed exclusions when varying the cross section by its
      theoretical uncertainty. The purple dashed thick (thin) line
      indicates the median (${\pm}1 \sigma$) expected exclusion. No
      interpretation is provided for the kinematic region defined by
      $0 < \mparent - \mlsp < 175\GeV$ or $\mparent < 300\GeV$, as
      discussed in the text. (Colour figure online.) \label{fig:limits-sms} }
  \end{center}
\end{figure*}

Figure~\ref{fig:limits-sms} shows the observed upper limit on the
production cross section at 95\% confidence level (CL) as a function
of the parent sparticle and LSP masses for various simplified
models. The point-to-point fluctuations are due to the finite number
of pseudo-experiments used to determine the observed upper limit.
The observed excluded regions are determined with NLO+NLL cross
sections for squark pair production assuming decoupled gluinos (and
vice versa), \ie the decoupled sparticle has a sufficiently high mass
such that it does not contribute significantly to the cross section.
Also shown are the observed excluded regions when varying the
production cross section by its theoretical uncertainty, and the
expected excluded region with the ${\pm}1$ standard-deviation
variations.

Two sets of excluded regions are provided for the model \Ttwo, as
shown in Figure~\ref{fig:limits-sms} (top left). The larger of the two
excluded regions is determined assuming an eightfold degeneracy for
the masses of the first- and second-generation squarks, \sQuaL and
\sQuaR (\sQua = \sU, \sD, \sS, and \sC), and decoupled
third-generation squarks and gluinos. The smaller of the two excluded
regions assumes the pair production of a single light squark, \suL,
with the gluino and all other squarks decoupled to high masses. The
models \TtwoBB and \TtwoTT assume the pair production of a single
bottom and top squark, respectively.

Table~\ref{tab:sms-efficiencies} lists the expected signal yields and
analysis efficiencies in the region $\scalht > 375\GeV$ for each of
the reference models defined in Table~\ref{tab:simplified-models}. The
yields and efficiencies are summed over the individual event
categories used for each interpretation, as listed in
Table~\ref{tab:categories}. The observed and expected upper limits
(95\% CL) on the cross section are also quoted, which can be compared
with the NLO+NLL SUSY production cross section and its theoretical
uncertainty.

\begin{table*}[tbhp]
  \topcaption{Summary of expected yields, analysis efficiencies, and
    upper limits for the reference models defined in
    Table~\ref{tab:simplified-models} using the event categories
    defined in Table~\ref{tab:categories}. The first row
    specifies the reference model. The second and third rows quote the
    expected yield and analysis efficiency (with statistical
    uncertainties) for the region $\scalht > 375\GeV$. The fourth
    row quotes the NLO+NLL SUSY production cross section (with
    theoretical uncertainty). For the model \Ttwo, this cross section
    assumes an eightfold mass degeneracy. In the case of only a
    single light squark, the cross section is $25 \pm
    4\unit{fb}$. The fifth and sixth rows quote the observed and
    expected upper limits (95\% CL) on the production cross section.
  }
  \label{tab:sms-efficiencies}
  \footnotesize
  \centering
  \newcommand{\mc}[1]{\multicolumn{2}{c}{#1}}
  \begin{tabular}{ lr@{$\,\pm\,$}lr@{$\,\pm\,$}lr@{$\,\pm\,$}lr@{$\,\pm\,$}lr@{$\,\pm\,$}lr@{$\,\pm\,$}l }
    \hline
    Reference model                 & \mc{\Ttwo} & \mc{\TtwoBB} & \mc{\TtwoTT} & \mc{\Tone} & \mc{\ToneBBBB} & \mc{\ToneTTTT}                             \\
    \hline
    Expected yield                  & 358.3      & 8.9          & 78.1         & 2.4        & 90.6           & 2.4 & 416  & 13  & 52.0 & 1.7 & 25.3 & 0.7 \\
    Efficiency [\%]                 & 16.0       & 0.4          & 10.2         & 0.3        & 2.9            & 0.1 & 10.4 & 0.3 & 9.4  & 0.3 & 2.9  & 0.1 \\
    Theoretical cross section [fb]  & 196        & 35           & 86           & 13         & 357            & 51  & 434  & 81  & 60   & 14  & 97   & 21  \\
    Observed upper limit [fb]       & \mc{113.2} & \mc{42.3}    & \mc{360.8}   & \mc{103.0} & \mc{15.0}      & \mc{46.2}                                  \\
    Expected upper limit [fb]       & \mc{103.1} & \mc{31.2}    & \mc{240.6}   & \mc{65.2}  & \mc{12.3}      & \mc{35.3}                                  \\
    \hline
  \end{tabular}
\end{table*}

The estimates of mass limits are determined from the observed
exclusion based on the theoretical production cross section, less
one-standard-deviation uncertainty. The most stringent mass limit on
the parent sparticle, $m_{\text{parent}}^{\text{best}}$, is generally
obtained at low LSP masses. Generally speaking, the excluded mass
range for \mparent is bounded from below by the kinematic region
considered for each model, yielding an exclusion that is generally
valid for the region $\mlsp + 175\gev \lesssim \mparent \lesssim
m_{\text{parent}}^{\text{best}}$. Whether an exclusion can be
determined for very small mass splittings, satisfying $\mparent -
\mlsp < 175\GeV$, requires further detailed studies of the modelling
of, for example, initial-state radiation, JES, or the identification
of b-quark jets. The upper bound on \mparent weakens for increasing
values of LSP mass until a value $m_{\mathrm{LSP}}^{\text{best}}$ is
reached, beyond which no exclusion on \mparent can be set.

Table~\ref{tab:sms-reach} summarises the most stringent observed and
expected mass limits, in terms of $m_{\text{parent}}^{\text{best}}$
and $m_{\mathrm{LSP}}^{\text{best}}$, obtained for the simplified
models considered in this paper. The observed exclusion for each
simplified model is generally weaker than expected at the level of
1--2 standard deviations. This feature is attributed to the small
upward fluctuations in data in either the region $\scalht > 875\GeV$
for the $\nb = 0$ category or $475 < \scalht < 675\GeV$ for the
categories of events satisfying $1 \leq \nb \leq 2$. Candidate events
in these regions have been examined and do not exhibit any
non-physical behaviour. The expected search sensitivity has improved
with respect to the analysis based on the $\sqrt{s} = 7\TeV$
dataset~\cite{RA1Paper2011FULL} by as much as 225 and 150\GeV for
$m_{\text{parent}}^{\textrm{best}}$ and
$m_{\textrm{LSP}}^{\textrm{best}}$, respectively.

\begin{table*}[tbhp]
  \topcaption{
    Summary of the mass limits obtained for various simplified
    models. The limits indicate the observed (expected) search
    sensitivity for each simplified model, where
    $m_{\text{parent}}^{\text{best}}$ and
    $m_{\mathrm{LSP}}^{\text{best}}$ represent the largest mass
    beyond which no limit can be set for squarks or gluinos and the
    LSP, respectively. Limits are quoted for the model \Ttwo assuming
    both an eightfold mass degeneracy ($\sQua$) and only a single
    light squark (\suL). No exclusion is observed in the mass
    parameter space considered for the model \TtwoTT.
  }
  \label{tab:sms-reach}
  \footnotesize
  \centering
  \begin{tabular}{ lccccccc }
    \hline
    Model                                 & \Ttwo ($\sQua$) & \Ttwo ($\suL$) & \TtwoBB   & \TtwoTT  & \Tone      & \ToneBBBB   & \ToneTTTT  \\
    \hline
    $\mparent^{\textrm{best}}$ [\GeVns{}] & 750 (850)       & 450 (475)      & 600 (675) & -- (520) & 950 (1050) & 1125 (1200) & 950 (1075) \\
    $\mlsp^{\textrm{best}}$ [\GeVns{}]    & 300 (325)       & 100 (125)      & 200 (250) & -- (100) & 450 (550)  & 650 (700)   & 325 (375)  \\
    \hline
  \end{tabular}
\end{table*}

\begin{figure*}[tbhp]
  \begin{center}
    \includegraphics[width=0.45\textwidth]{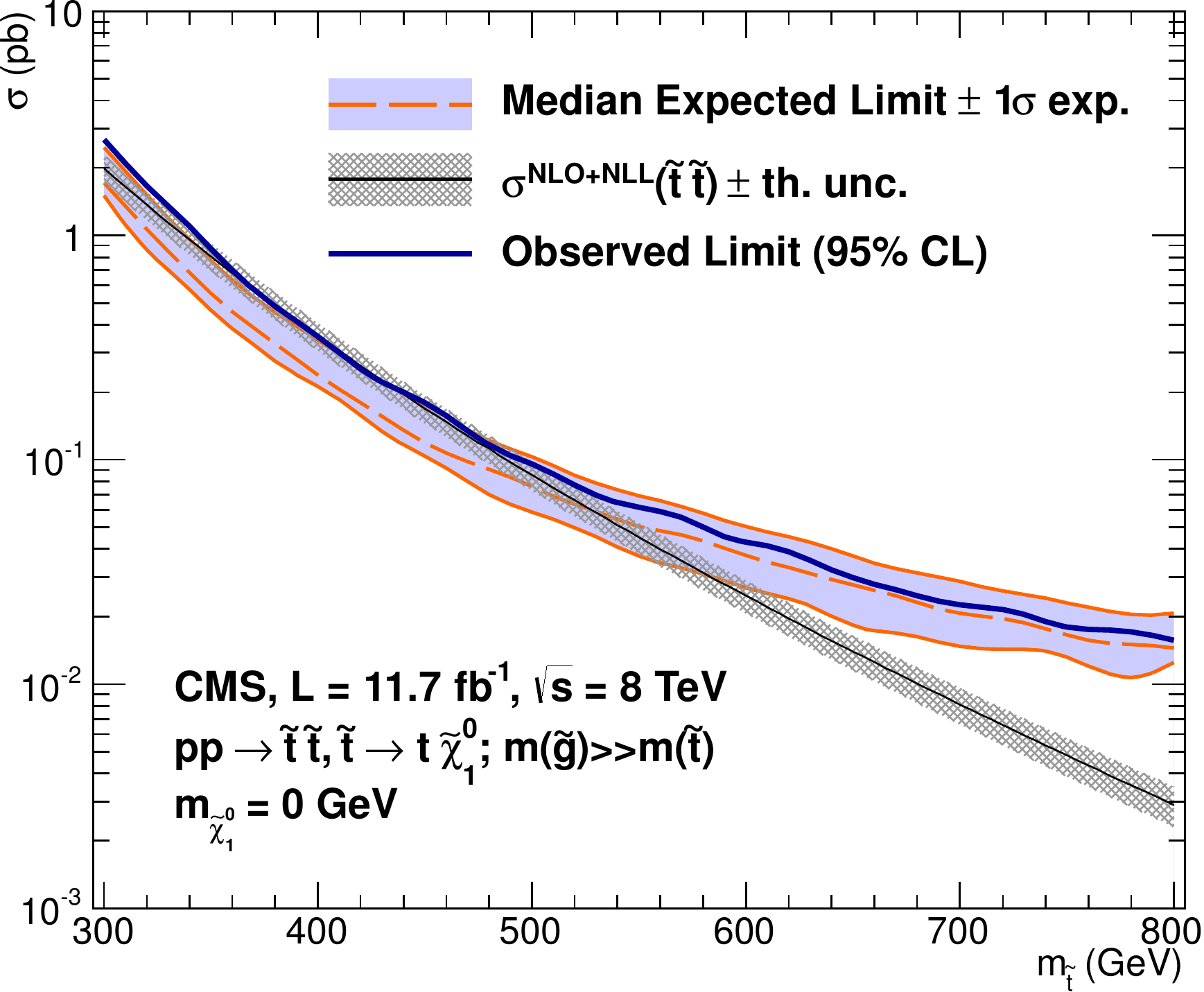} \,
    \includegraphics[width=0.45\textwidth]{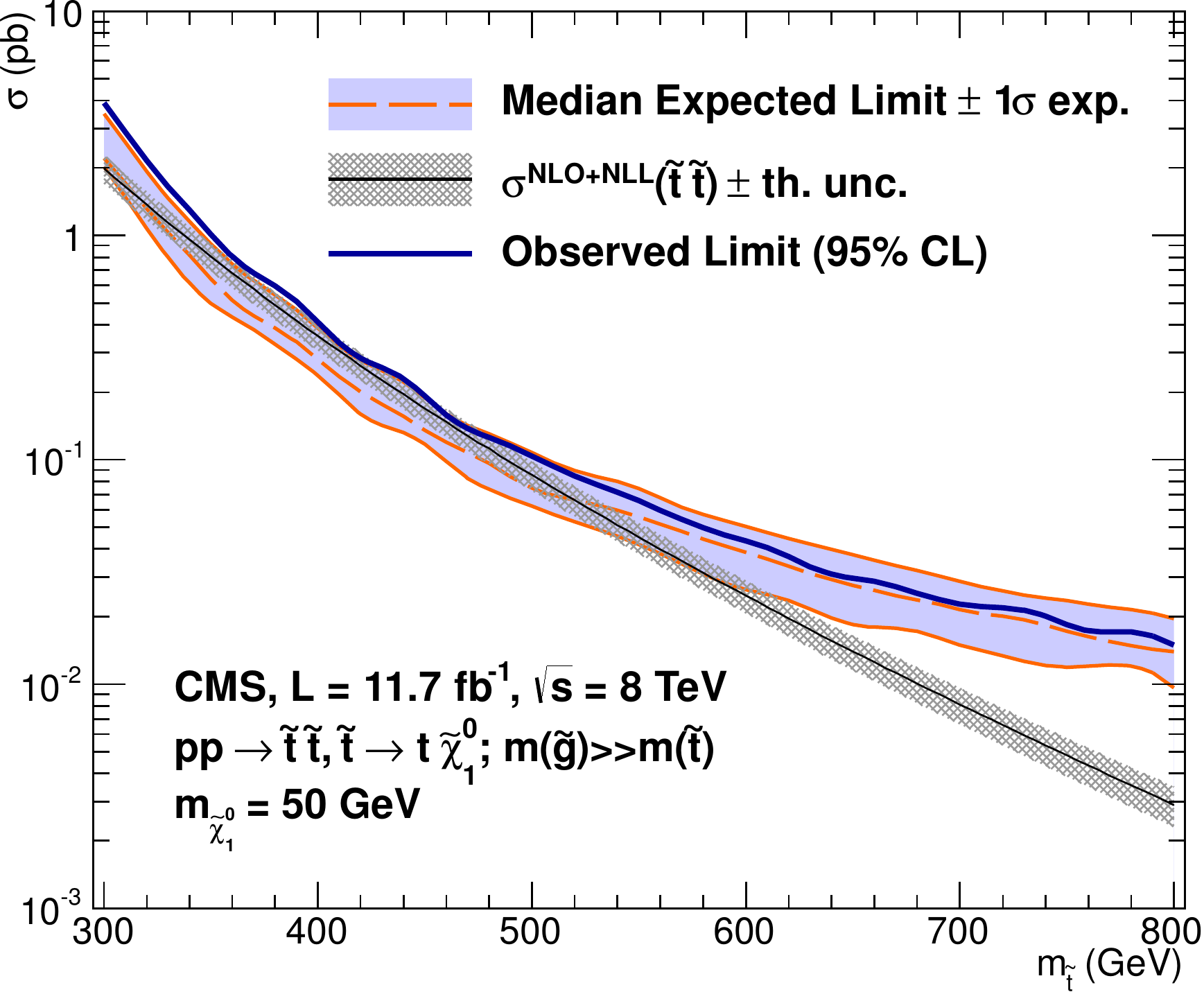}  \\
    \includegraphics[width=0.45\textwidth]{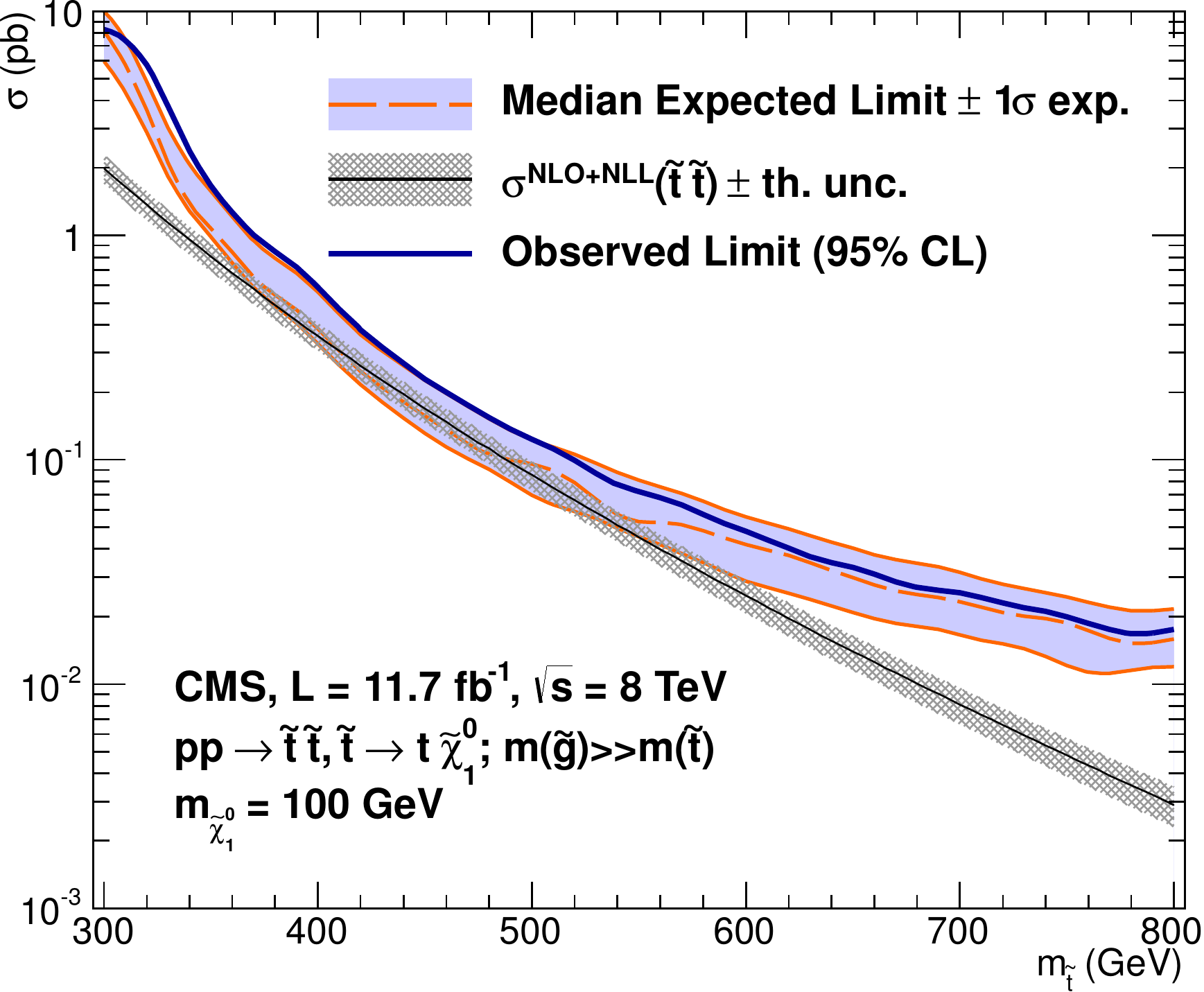} \,
    \includegraphics[width=0.45\textwidth]{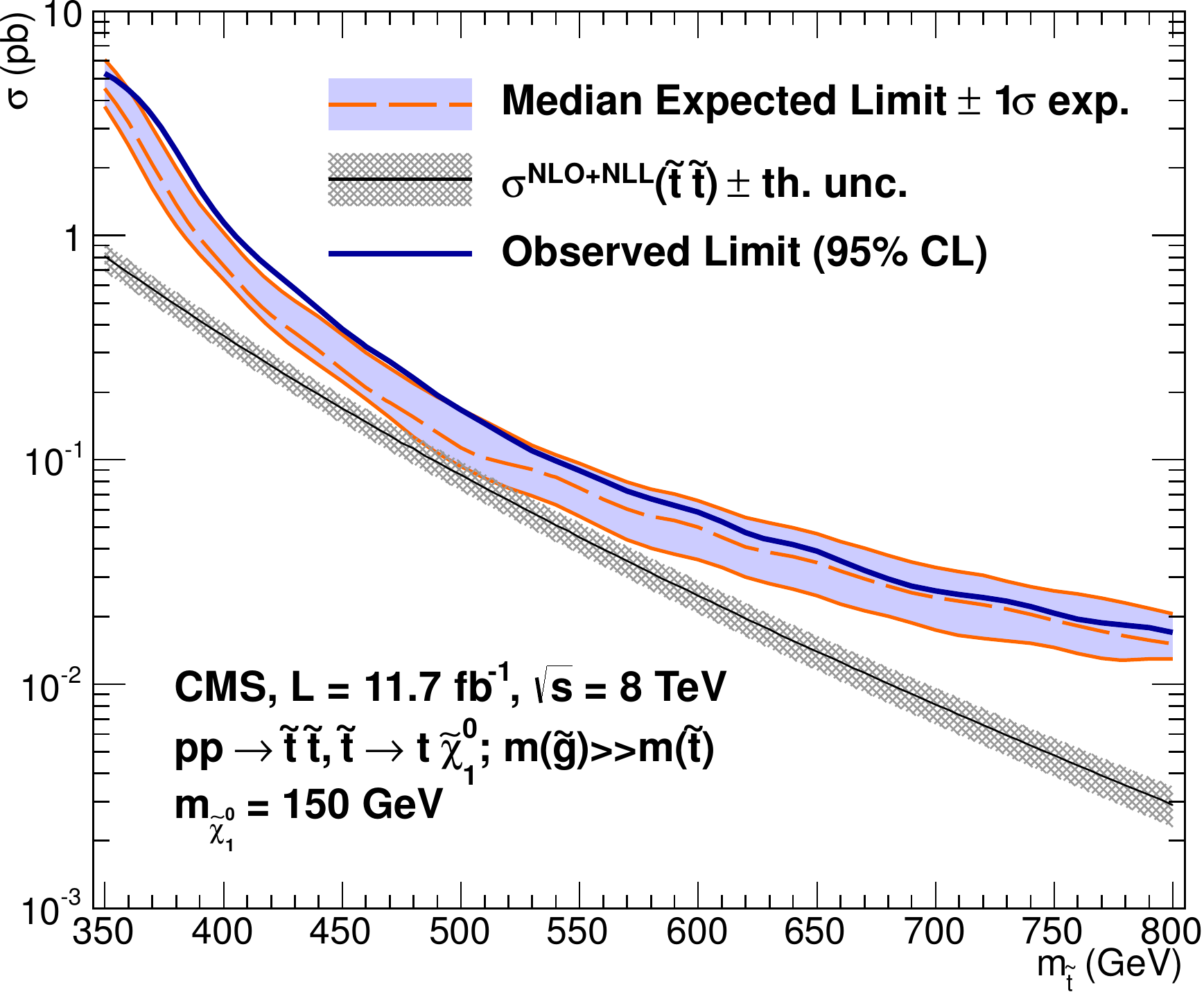} \\
    \caption{\label{fig:t2tt-1d} Excluded cross sections versus
      top-squark mass $m_{\sTop}$ for the model \TtwoTT, in which
      pair-produced top squarks each decay to a top quark and the LSP
      with a mass $m_\mathrm{LSP} = 0$ (top left), 50 (top right), 100
      (bottom left), and 150\gev (bottom right). The observed upper
      limit (95\% CL) on the production cross section is shown as a
      function of $m_{\sTop}$ (solid line), along with the expected
      upper limit and $\pm1\sigma$ experimental uncertainties
      (long-dashed line with shaded band), and the NLO+NLL top-squark
      pair production cross section and theoretical uncertainties
      (dotted line with shaded band). (Colour figure online.)}
  \end{center}
\end{figure*}

Figure~\ref{fig:t2tt-1d} shows the observed upper limit at 95\% CL on
the production cross section as a function of the top-squark mass
($m_{\sTop}$) for the model \TtwoTT when considering different LSP
masses in the range 0--150\GeV. No exclusion on possible top-squark
masses is observed when considering the theoretical production cross
section, less $1\sigma$ uncertainty. However, the expected exclusion
covers the ranges 300--520, 320--520, and 420-480\GeV for $\mlsp = 0$,
50, and 100\GeV, respectively. No exclusion is expected for the LSP
with a mass greater than 100\GeV. The expected reach for the \TtwoTT
model is summarised in Table~\ref{tab:sms-reach}.

\section{Summary}

An inclusive search for supersymmetry with the CMS experiment is
reported, based on a data sample of pp collisions collected at
$\sqrt{s} = 8\TeV$, corresponding to an integrated luminosity of $11.7
\pm 0.5\fbinv$.  Final states with two or more energetic jets and
significant \met, as expected from the production and decay of massive
squarks and gluinos, have been analysed.

The analysis strategy is to maximise the sensitivity of the search to
a wide variety of SUSY event topologies arising from squark-squark,
squark-gluino, and gluino-gluino production and decay, particularly
those with third-generation squark signatures, while still maintaining
the inclusive nature of the search.  The signal region is binned
according to the number of reconstructed jets, the scalar sum of the
transverse energy of jets, and the number of jets identified to
originate from bottom quarks. The sum of standard model backgrounds
per bin has been estimated from a simultaneous binned likelihood fit
to event yields in the signal region and $\mu + \text{jets}$, $\mu\mu
+ \text{jets}$, and $\gamma + \text{jets}$ control samples. The
observed yields in the signal region are found to be in agreement with
the expected contributions from standard model processes. Limits are
set in the SUSY particle mass parameter space of simplified models,
with an emphasis on the different production mechanisms of coloured
SUSY particles, third-generation squark signatures, and
compressed-spectrum scenarios. The results can also be used to perform
interpretations in other relevant models, such as the CMSSM.

In the context of simplified models, gluino masses below ${\sim}1\TeV$
are excluded at the 95\% CL under the assumptions that gluinos are
produced in pairs and each decays to a quark-antiquark pair and a
light LSP via an off-shell squark. The mass limit varies in the range
950--1125\GeV depending on the squark flavour. The most constraining
mass limits on the 
LSP from the decay of a gluino are in the range 325--650\GeV depending
on the decay mode.
For the direct production of first- and second-generation squark
pairs, each of which is assumed to decay to a quark of the same
flavour and the LSP, masses below 750\GeV are excluded (95\% CL) under
the assumption of an eightfold mass-degeneracy. The most constraining
mass limit on the LSP is 300\GeV. These limits weaken to 450 and
100\GeV respectively if only a single squark is assumed to be
light. For the direct production of bottom squark pairs, each of which
is assumed to decay to a bottom quark and the LSP, masses below
600\GeV are excluded. No exclusion is possible for an LSP mass beyond
200\GeV. No exclusion is observed for the direct pair production of
top squarks, each of which is assumed to decay to a top quark and the
LSP. However, an exclusion is expected for top squark masses as high
as $\sim$500\GeV and an LSP mass as high as 100\GeV. The limits on the
LSP masses are also generally valid for compressed-spectrum models
with mass splittings between the parent sparticle and LSP as low as
$\sim$200\GeV.

The analysis strategy reported here, in conjunction with the increase
in centre-of-mass energy to 8\TeV, has increased the coverage of the
SUSY parameter space with respect to previous searches.  However, a
large range of the SUSY parameter space still remains to be probed by
the LHC.
\section*{Acknowledgements}

{\tolerance=1200\hyphenation{Bundes-ministerium Forschungs-gemeinschaft Forschungs-zentren} We congratulate our colleagues in the CERN accelerator departments for the excellent performance of the LHC and thank the technical and administrative staffs at CERN and at other CMS institutes for their contributions to the success of the CMS effort. In addition, we gratefully acknowledge the computing centres and personnel of the Worldwide LHC Computing Grid for delivering so effectively the computing infrastructure essential to our analyses. Finally, we acknowledge the enduring support for the construction and operation of the LHC and the CMS detector provided by the following funding agencies: the Austrian Federal Ministry of Science and Research and the Austrian Science Fund; the Belgian Fonds de la Recherche Scientifique, and Fonds voor Wetenschappelijk Onderzoek; the Brazilian Funding Agencies (CNPq, CAPES, FAPERJ, and FAPESP); the Bulgarian Ministry of Education, Youth and Science; CERN; the Chinese Academy of Sciences, Ministry of Science and Technology, and National Natural Science Foundation of China; the Colombian Funding Agency (COLCIENCIAS); the Croatian Ministry of Science, Education and Sport; the Research Promotion Foundation, Cyprus; the Ministry of Education and Research, Recurrent financing contract SF0690030s09 and European Regional Development Fund, Estonia; the Academy of Finland, Finnish Ministry of Education and Culture, and Helsinki Institute of Physics; the Institut National de Physique Nucl\'eaire et de Physique des Particules~/~CNRS, and Commissariat \`a l'\'Energie Atomique et aux \'Energies Alternatives~/~CEA, France; the Bundesministerium f\"ur Bildung und Forschung, Deutsche Forschungsgemeinschaft, and Helmholtz-Gemeinschaft Deutscher Forschungszentren, Germany; the General Secretariat for Research and Technology, Greece; the National Scientific Research Foundation, and National Office for Research and Technology, Hungary; the Department of Atomic Energy and the Department of Science and Technology, India; the Institute for Studies in Theoretical Physics and Mathematics, Iran; the Science Foundation, Ireland; the Istituto Nazionale di Fisica Nucleare, Italy; the Korean Ministry of Education, Science and Technology and the World Class University program of NRF, Republic of Korea; the Lithuanian Academy of Sciences; the Mexican Funding Agencies (CINVESTAV, CONACYT, SEP, and UASLP-FAI); the Ministry of Science and Innovation, New Zealand; the Pakistan Atomic Energy Commission; the Ministry of Science and Higher Education and the National Science Centre, Poland; the Funda\c{c}\~ao para a Ci\^encia e a Tecnologia, Portugal; JINR (Armenia, Belarus, Georgia, Ukraine, Uzbekistan); the Ministry of Education and Science of the Russian Federation, the Federal Agency of Atomic Energy of the Russian Federation, Russian Academy of Sciences, and the Russian Foundation for Basic Research; the Ministry of Science and Technological Development of Serbia; the Secretar\'{\i}a de Estado de Investigaci\'on, Desarrollo e Innovaci\'on and Programa Consolider-Ingenio 2010, Spain; the Swiss Funding Agencies (ETH Board, ETH Zurich, PSI, SNF, UniZH, Canton Zurich, and SER); the National Science Council, Taipei; the Thailand Center of Excellence in Physics, the Institute for the Promotion of Teaching Science and Technology of Thailand and the National Science and Technology Development Agency of Thailand; the Scientific and Technical Research Council of Turkey, and Turkish Atomic Energy Authority; the Science and Technology Facilities Council, UK; the US Department of Energy, and the US National Science Foundation.
Individuals have received support from the Marie-Curie programme and the European Research Council and EPLANET (European Union); the Leventis Foundation; the A. P. Sloan Foundation; the Alexander von Humboldt Foundation; the Belgian Federal Science Policy Office; the Fonds pour la Formation \`a la Recherche dans l'Industrie et dans l'Agriculture (FRIA-Belgium); the Agentschap voor Innovatie door Wetenschap en Technologie (IWT-Belgium); the Ministry of Education, Youth and Sports (MEYS) of Czech Republic; the Council of Science and Industrial Research, India; the Compagnia di San Paolo (Torino); and the HOMING PLUS programme of Foundation for Polish Science, cofinanced from European Union, Regional Development Fund.\par}
\clearpage
\bibliography{auto_generated}
\cleardoublepage \appendix\section{The CMS Collaboration \label{app:collab}}\begin{sloppypar}\hyphenpenalty=5000\widowpenalty=500\clubpenalty=5000\textbf{Yerevan Physics Institute,  Yerevan,  Armenia}\\*[0pt]
S.~Chatrchyan, V.~Khachatryan, A.M.~Sirunyan, A.~Tumasyan
\vskip\cmsinstskip
\textbf{Institut f\"{u}r Hochenergiephysik der OeAW,  Wien,  Austria}\\*[0pt]
W.~Adam, T.~Bergauer, M.~Dragicevic, J.~Er\"{o}, C.~Fabjan\cmsAuthorMark{1}, M.~Friedl, R.~Fr\"{u}hwirth\cmsAuthorMark{1}, V.M.~Ghete, N.~H\"{o}rmann, J.~Hrubec, M.~Jeitler\cmsAuthorMark{1}, W.~Kiesenhofer, V.~Kn\"{u}nz, M.~Krammer\cmsAuthorMark{1}, I.~Kr\"{a}tschmer, D.~Liko, I.~Mikulec, D.~Rabady\cmsAuthorMark{2}, B.~Rahbaran, C.~Rohringer, H.~Rohringer, R.~Sch\"{o}fbeck, J.~Strauss, A.~Taurok, W.~Treberer-treberspurg, W.~Waltenberger, C.-E.~Wulz\cmsAuthorMark{1}
% ...
\vskip\cmsinstskip
% ...
1:~~Also at Vienna University of Technology, Vienna, Austria\\
2:~~Also at CERN, European Organization for Nuclear Research, Geneva, Switzerland\\
% ...

\end{sloppypar}
\end{document}